\documentclass[doublespacing]{bmcart}

\RequirePackage{hyperref}
\usepackage[utf8]{inputenc} 

\usepackage{graphicx}
\usepackage{dcolumn}
\usepackage{bm}
\usepackage{braket}
\usepackage{comment}
\usepackage{amssymb}
\usepackage{amsmath}
\usepackage{color}
\usepackage{textcomp}
\usepackage{booktabs}
\usepackage{subfigure}
\usepackage{siunitx}

\startlocaldefs
\endlocaldefs

\begin{document}

\begin{frontmatter}

\begin{fmbox}
\dochead{Tutorial}


\title{Light control with Weyl semimetals}


\author[
   addressref={aff1},                   
   noteref={n1,n2},                        
]{\inits{CG}\fnm{Cheng} \snm{Guo}}
\author[
   addressref={aff2,aff3},
    noteref={n1},
]{\inits{VsA}\fnm{Viktar S.} \snm{Asadchy}}
\author[
   addressref={aff4},
   noteref={n1},
]{\inits{BZ}\fnm{Bo} \snm{Zhao}}
\author[
   addressref={aff2},
    noteref={n3},
]{\inits{SF}\fnm{Shanhui} \snm{Fan}}


\address[id=aff1]{
  \orgname{Department of Applied Physics}, 
  \street{Stanford University},                     %
  \postcode{94305}                                
  \city{Stanford,  California},                              
  \cny{USA}                                    
}
\address[id=aff2]{%
  \orgname{Ginzton Laboratory and Department of Electrical Engineering},
  \street{Stanford University},
  \postcode{94305}
  \city{Stanford,  California},
  \cny{USA}
}
\address[id=aff3]{%
  \orgname{Department of Electronics and Nanoengineering},
  \street{Aalto University},
  \postcode{02150}
  \city{Espoo},
  \cny{Finland}
}
\address[id=aff4]{%
  \orgname{Department of Mechanical Engineering},
  \street{University of Houston},
  \postcode{77204}
  \city{Houston, Texas},
  \cny{USA}
}


\begin{artnotes}

\note[id=n1]{Equal contributor\\} 

\note[id=n2]{Correspondence: guocheng@stanford.edu\\}

\note[id=n3]{Correspondence: shanhui@stanford.edu}
\end{artnotes}

\end{fmbox}


\begin{abstractbox}
\begin{abstract} 
Weyl semimetals are topological materials whose electron quasiparticles obey the Weyl equation. They possess many unusual properties that may lead to new applications. This is a tutorial review of the optical properties and applications of Weyl semimetals. We review the basic concepts and optical responses of Weyl semimetals, and survey their applications in optics and thermal photonics. We hope this pedagogical text will motivate further research on this emerging topic. 

\end{abstract}




\end{abstractbox}
%

\end{frontmatter}




\section*{Introduction}

Weyl semimetals are topological materials whose low-energy excitations obey the Weyl equation~\cite{Herring1937,armitage2018a}. In a Weyl semimetal, the conduction and valence bands touch at discrete points in momentum space, called Weyl nodes. Weyl nodes are monopoles of the Berry curvature and are robust under generic perturbations. The quasiparticles near the Weyl nodes are analogous to Weyl fermions in high-energy physics~\cite{srednicki2007}; they exhibit linear dispersion and well-defined chirality. 

The nontrivial topology of Weyl semimetals leads to many unusual electronic, magnetic, thermal, and optical properties~\cite{armitage2018a,gorbar2021,nagaosa2020}. These intriguing features have been extensively studied in the literature. Besides these fundamental interests, Weyl semimetals may also enable new opportunities in practical applications. For example, photonic applications include compact optical isolators and circulators~\cite{kotov_giant_2018,asadchy_sub-wavelength_2020,park_violating_2021}, orbital angular momentum detectors~\cite{ji_photocurrent_2020,lai_direct_2022}, higher-order harmonic generation~\cite{wu_giant_2017,almutairi_four-wave_2020,cheng_third-order_2020}, and nonreciprocal thermal emitters~\cite{zhao_axion-field-enabled_2020,ChenWeyl1,ChenWeyl2} among many others. However, such an application-oriented exploration is still at an early stage, which requires more joint efforts from scientists and engineers.

This text is a tutorial review of Weyl semimetals that should be of interest to researchers working in photonics, applied physics, and optical engineering. We start with the basic concepts of semimetals and Weyl semimetals (Sec.~\ref{sec:semimetals_and_Weyl_semimetals}). Then, we review the formalism of axion electrodynamics and derive the optical responses of Weyl semimetals (Sec.~\ref{sec:optical_properties}). Next, we survey the broad potential applications of Weyl semimetals in optics (Sec.~\ref{sec:dev}). Finally, we discuss the applications of Weyl semimetals in thermal photonics (Sec.~\ref{sec:thermal_dev}). We hope that our survey will motivate further exploration of photonic applications with Weyl semimetals.

\section{Semimetals and Weyl semimetals}\label{sec:semimetals_and_Weyl_semimetals}

Weyl semimetals are a special class of semimetals. They exhibit common properties of semimetals as well as some  unique characteristics. This section provides a brief introduction to semimetals and Weyl semimetals. In Sec.~\ref{subsec:semimetals}, we review the basic concept and common properties of semimetals. We also discuss typical behaviors and material examples of conventional semimetals. In Sec.~\ref{subsec:Weyl_semimetals}, we review the basic concept and physical properties of Weyl semimetals. 

\subsection{Semimetals}\label{subsec:semimetals}

\begin{figure}[htbp]
    \centering
    \includegraphics[width=0.7\textwidth]{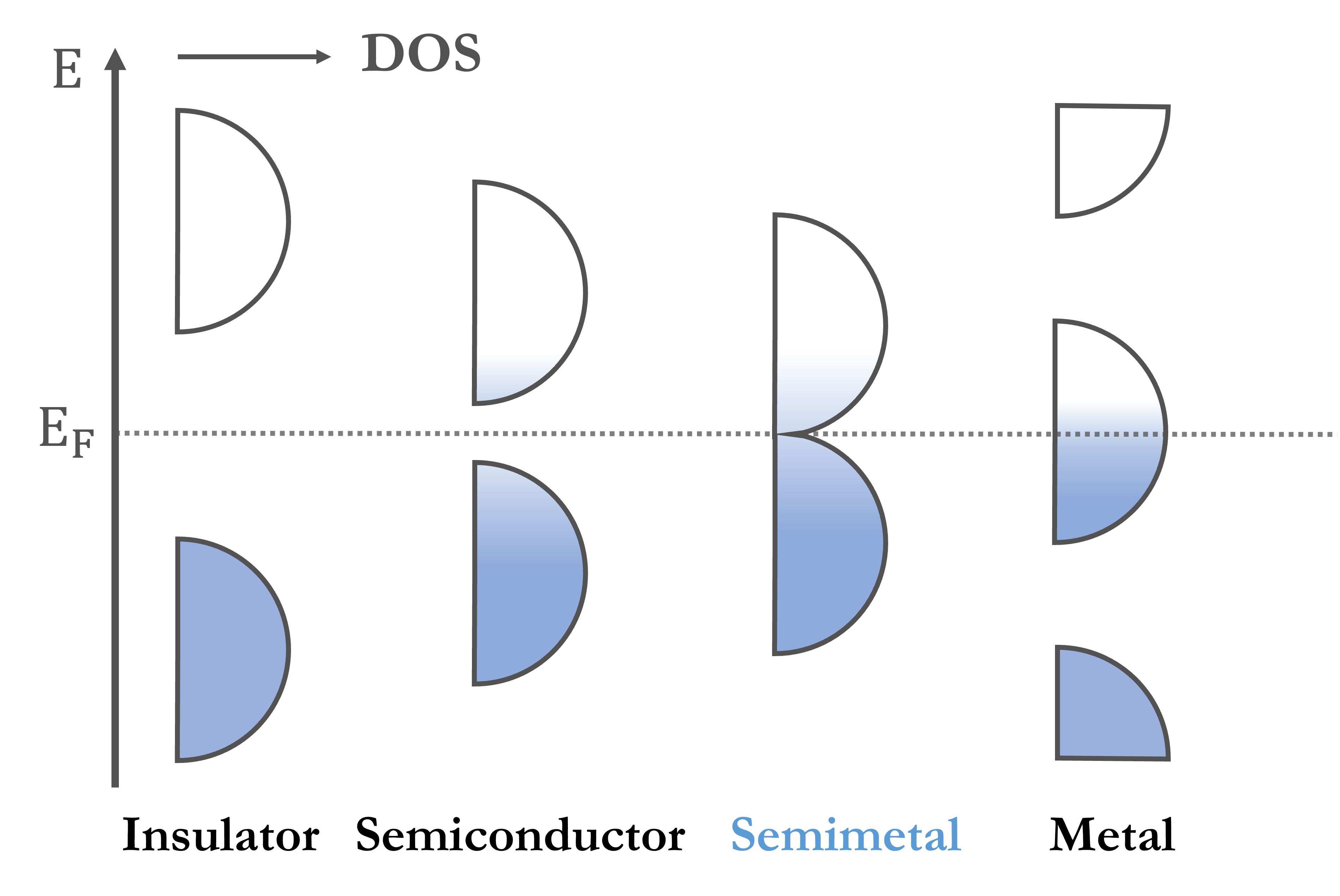}
    \caption{\csentence{Classification of solids.} Schematic density of states (DOS) of insulators, semiconductors, semimetals, and metals. $E_F$ denotes the Fermi level. }
    \label{fig:DOS}
\end{figure}

According to the band theory~\cite{ashcroft_solid_1976,burns1990}, solids can be classified as insulators, semiconductors, semimetals, and metals  (Fig.~\ref{fig:DOS}). An insulator or a semiconductor has a band gap between the valence and conduction bands; the band gap is larger for an insulator (\SI{>4}{eV}) than for a semiconductor (\SI{<4}{eV}). A \emph{semimetal} has a very small overlap between the conduction and valence bands and a negligible density of states at the Fermi level. A metal has a partially filled conduction band and an appreciable density of states at the Fermi level.

Different band structures lead to different physical properties. For example, the carrier concentration is \SI{>e22}{cm^{-3}} for normal metals, \SIrange[range-phrase = --, range-units = single]{\sim e17}{e20}{cm^{-3}} for semimetals, and \SIrange[range-phrase = --, range-units = single]{\sim e6}{e13}{cm^{-3}} for intrinsic semiconductors such as GaAs, Si, and Ge. Consequently, the electrical conductivities are \SIrange[range-phrase = --, range-units = single]{\sim e5}{e6}{\per \ohm \per \cm} for normal metals, \SI{\sim e4}{\per \ohm \per \cm} for semimetals, and \SIrange[range-phrase = --, range-units = single]{\sim e-8}{e-1}{\per \ohm \per \cm} for intrinsic semiconductors~\cite{burns1990}. 

Semimetals, the focus of our study, are probably the least known among the four types. An intrinsic semimetal has an equal number of electrons and holes. Like a normal metal, its conductivity increases as the temperature is lowered. Like a semiconductor, it can be doped with proper impurities to vary the number of electrons and holes. Its electronic properties are also sensitive to pressure since pressure changes the internuclear distances, which sensitively changes the amount of band overlap and causes large changes in the carrier concentrations.  Besides low carrier concentrations, semimetals typically have small effective masses and high dielectric constants. Nonmagnetic semimetals typically exhibit high diamagnetic susceptibilities and huge electron $g$-values~\cite{burns1990}.

\begin{figure}[htbp]
    \centering
    \includegraphics[width=0.7\textwidth]{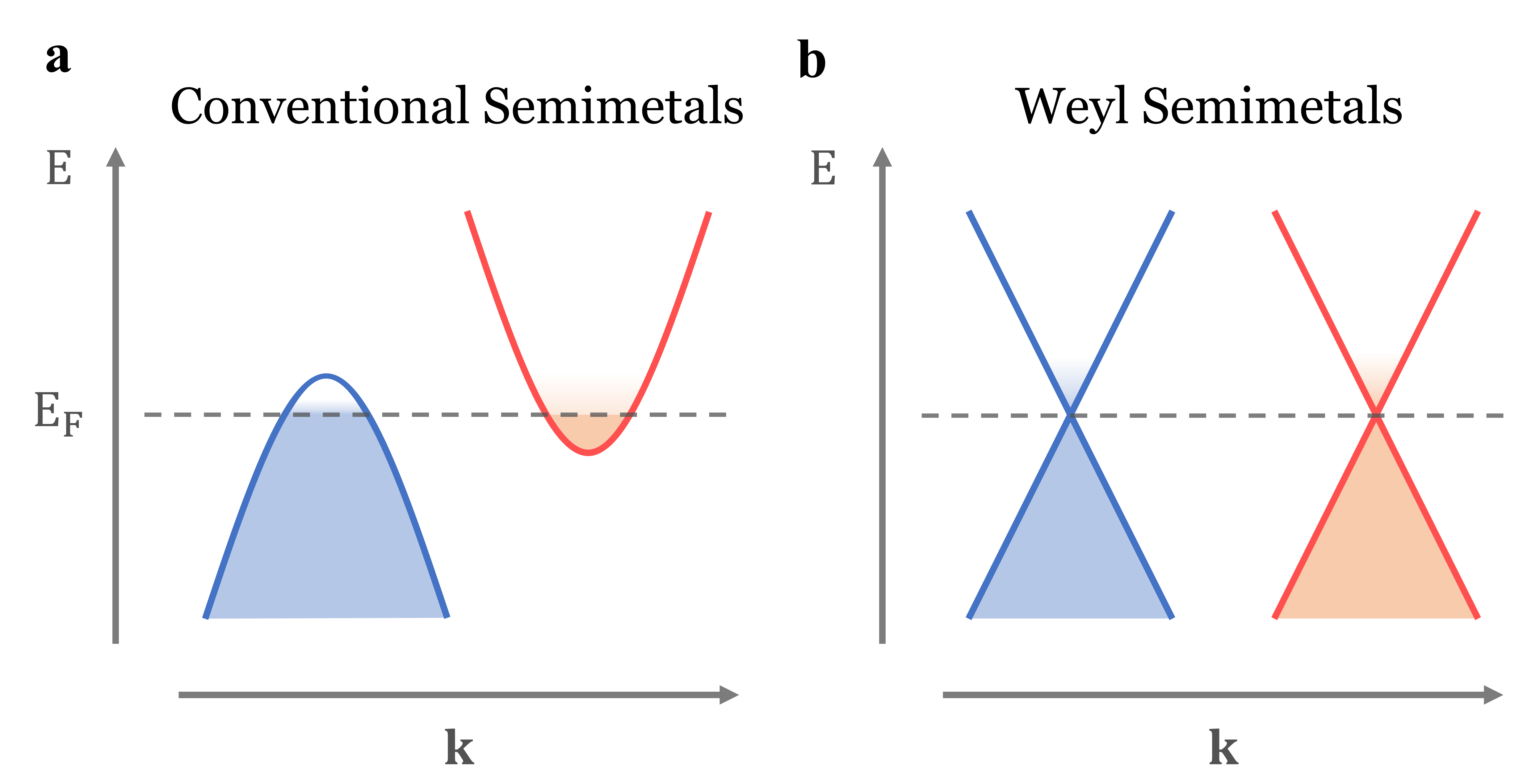}
    \caption{\csentence{Conventional versus Weyl semimetals.} Schematic band structure for (a) conventional semimetals,  (b) Weyl semimetals.}
    \label{fig:classic_vs_weyl}
\end{figure}

A conventional semimetal is a ``semiconductor" with a negative indirect bandgap (Fig.~\ref{fig:classic_vs_weyl}a). In an intrinsic semimetal, the bottom of the conduction band is slightly lower than the top of the valence band, and the Fermi level sits in between. Hence, the semimetal has charge carries of both types (holes and electrons). Typically, the pockets of electrons and holes are located at different positions in the wavevector space~\cite{burns1990}.

Classic examples of semimetals include the group 5A elements: arsenic, antimony, and bismuth~\cite{burns1990}. They have five valence electrons per atom, so if there were one atom per primitive cell the material would be a metal. However, there are two atoms per primitive cell. With the ten valence electrons, the crystal could be an insulator but there is a small overlap in energy between the conduction and valence bands, resulting in semimetal behavior~\cite{golin1968}.

Another well-known semimetal is graphite. It consists of stacked layers of graphene. Since the interlayer bonding is very weak, the band overlap is tiny~\cite{slonczewski1958}. Hence, graphite is a semimetal, and the Fermi surface consists primarily of tiny pockets of electrons and holes at different wavevectors, with carrier densities of around $n_{e}= n_{h} = \SI{3e18}{cm^{-3}}$~\cite{ashcroft_solid_1976}.

\subsection{Weyl semimetals}\label{subsec:Weyl_semimetals}

Weyl semimetals are a unique class of semimetals. Unlike conventional semimetals, in Weyl semimetals, the valence and conduction bands touch at discrete points in the wavevector space. Near the touching point, the band dispersion is linear (Fig.~\ref{fig:classic_vs_weyl}b). 

Weyl semimetals are named after Hermann Weyl (1885-1955). In 1929, Weyl proposed the Weyl equation~\cite{weyl1929}, which simplifies the Dirac equation for a \emph{massless} relativistic spin $\frac{1}{2}$ particle:
\begin{equation}\label{eq:Weyl_equation}
i \hbar \frac{\partial \Psi}{\partial t} = \hat{H} \Psi = \pm c \bm{p}\cdot \bm{\sigma} \Psi.
\end{equation}
Here $\hbar$ is the reduced Planck constant, $c$ is the speed of light, $\bm{p}= (p_{x}, p_{y}, p_{z})$ is the momentum operator, $\bm{\sigma}=(\sigma_{x},\sigma_{y},\sigma_{z})$ is a vector whose components are the Pauli matrices, and $\Psi$ is a two-component field called a Weyl spinor~\cite{srednicki2007}. The Weyl equation describes a massless particle with linear dispersion, called a Weyl fermion. The $\pm$ sign reflects the existence of two types of Weyl fermions: right-handed ($+$) and left-handed ($-$). Hence, unlike the Dirac equation, the Weyl equation violates parity.

Physicists have been searching for Weyl fermions in nature. However, none of the observed elementary particles are Weyl fermions~\cite{armitage2018a}. It was thought that neutrinos could be Weyl fermions. However, the discovery of neutrino oscillation in 1998~\cite{super-kamiokandecollaboration1998} shows that neutrinos are massive and thus cannot be Weyl fermions.

Despite their absence in high-energy physics, Weyl fermions may be realized in condensed-matter systems. In 1937, Conyers Herring proposed the concept of Weyl semimetals~\cite{Herring1937}. He realized that when two electronic bands cross accidentally, the generic band dispersions near a touching point are linear in all directions. In the simplest scenario, the effective Hamiltonian reads~\cite{vishwanath2015}:
\begin{equation}\label{eq:Hamiltonian_Herring}
\hat{H} = \pm v_F \bm{p}\cdot \bm{\sigma} 
\end{equation}
where $\bm{p}$ is the momentum deviation from the touching point, and $v_{F}$ is the Fermi velocity. The resemblance between Eqs.~(\ref{eq:Weyl_equation}) and (\ref{eq:Hamiltonian_Herring}) is evident. As a result, these touching points are called Weyl nodes, and the quasiparticles near them are reminiscent of Weyl fermions. Each Weyl point has a definite chirality, either right-handed ($+$) or left-handed ($-$). For crystals where the wavevector space is the Brillouin zone and hence is compact, the Nielsen-Ninomiya no-go theorem asserts that left- and right-handed Weyl points always appear in pairs~\cite{nielsen1981,nielsen1981a,nielsen1981b,sun2018a}. Hence, it is impossible that only a single Weyl point exists in the momentum space for solid state systems.

Weyl points are robust to generic perturbations: The Hamiltonian remains gapless while Weyl nodes move around. This is because for a Hermitian matrix, having a pair of coalescing eigenvalues is an effect with codimension three; generically, one needs to tune three real parameters to realize such an effect. This fact, known as the von Neumann-Wigner theorem~\cite{vonneuman1929}, explains why generically the accidental degeneracies of two bands only occur at isolated points in the three-dimensional momentum space and why such degeneracies persist under perturbation. From a topological perspective, the robustness of Weyl points can be understood from Berry curvature~\cite{vanderbilt2018a}. Berry curvature arises from the variation of the periodic part of the Bloch wave function with respect to wavevetors, and is mathematically analogous to a magnetic field in momentum space. Weyl points are monopoles of Berry curvature with quantized charges, which are analogous to quantized magnetic monopoles for magnetic fields. Weyl points can only be destroyed when two Weyl nodes of opposite chiralities are moved together and annihilated with each other.

For a Weyl semimetal, the existence of Weyl nodes in the bulk band structure leads to the presence of Fermi arc surface states when the Weyl semimetal is truncated. Such surface states correspond to an open Fermi arc that connects the projection in the surface Brillouin zone of two Weyl nodes of opposite chiralities. The existence of such surface Fermi arcs is protected by the nontrivial chiral charge of the Weyl nodes~\cite{belopolski2016,xu2016d}. This is an example of  the general principle of bulk-boundary correspondence in topological materials.

Although the Weyl semimetal was theoretically proposed many decades ago, its experimental demonstration was quite recent. A material must satisfy some necessary conditions to be a potential candidate for Weyl semimetals~\cite{yan_topological_2017}. First, it must break either time-reversal or spatial inversion symmetry~\cite{armitage2018a}. In a system that satisfies both time-reversal and spatial inversion symmetry, all the bands must be doubly degenerate in the whole wavevector space due to the Kramers' degeneracy theorem~\cite{kramers1930}; this excludes the existence of Weyl nodes, which only appear as the accidental degeneracy of two non-degenerate bands. The minimum numbers of Weyl nodes are $2$ and $4$ for Weyl semimetals that break time-reversal symmetry and inversion symmetry, respectively. Second, the Weyl nodes must be located near the Fermi level, so that Weyl fermions may emerge as low-energy excitations.

Weyl semimetals were first discovered in 2015~\cite{xu2015,Lv2015} in non-centrosymmetric crystals, TaAs family (TaAs~\cite{xu2015,Lv2015}, TaP~\cite{xu2015b,zhang2017e}, NbAs~\cite{xu2015d}, and NbP~\cite{zheng2016b,zheng2017,Gooth2017}). Later, magnetic Weyl semimetals that break time-reversal symmetry have also been discovered~\cite{morali2019,belopolski2019,liu2019}. A number of other Weyl semimetals have since been discovered, as reviewed in Ref.~\cite{fisher2019}.

Weyl semimetals exhibit a rich variety of novel phenomena. Besides surface Fermi arc states, they also exhibit chiral anomaly~\cite{xiong2015,zhang2016g,ong2021}, unconventional charge and heat transport~\cite{zhang2015e,Gooth2017}, strain-induced axial gauge fields~\cite{weststrom_designer_2017,cortijo_elastic_2015,pikulin_chiral_2016}, novel collective modes~\cite{song2019b}, and so on. The properties of Weyl semimetals have been summarized in numerous reviews and monographs: For general overviews, see Refs.~\cite{vafek2014,armitage2018a,burkov2018a}; for electronic properties, see Ref.~\cite{gorbar2021}; for magnetic properties, see Refs.~\cite{witczak-krempa2014,smejkal2017,mikitik2019}; for transport properties, see Refs.~\cite{hosur2013a,burkov2015a,wang2017f,gorbar2018,wang2018g,hu2019b}; for topological properties, see Refs.~\cite{chiu2016a,witten2016}; for \emph{ab initio} calculation, see Ref.~\cite{weng2016}; for experimental studies, see Ref.~\cite{hasan2015,hasan2017a,yan2017,zheng2018}; for material properties, see Ref.~\cite{fisher2019}.
Complementary to these works, the present tutorial review will focus on the optical properties of Weyl semimetals.

\section{Optical properties of Weyl semimetals}\label{sec:optical_properties}

In this section, we derive the linear optical properties of Weyl semimetals. We consider the simplest Weyl semimetal with two nodes separated by $2\bm{b}$ in their wavevectors and $2 \hbar b_0$ in their energy (Fig.~\ref{fig:energy}a). The constitutive relations for such an ideal Weyl semimetal are:
\begin{equation}\label{eq:main_result}
\bm{D} = \varepsilon_D(\omega) \bm{E} + \frac{ie^2}{2\pi^2\hbar \omega}\left(-b_0\bm{B}+ \bm{b}\times \bm{E}\right), \qquad  \bm{H} = \frac{1}{\mu_0}\bm{B}.
\end{equation}
The derivation of Eq.~(\ref{eq:main_result}) is the main aim of this section. The expression of $\bm{D}$ has two terms; the first term is referred to as the Dirac term and the second the axion term. The Dirac term describes the permittivity of a corresponding Dirac semimetal (Fig.~\ref{fig:energy}b), i.e.,~a Weyl semimetal with two overlapping Weyl nodes ($\bm{b}=\bm{0}, b_0 = 0$). The axion term captures the effects of Weyl node separation. Its name is taken from axion electrodynamics. Its first term represents the chiral magnetic effect, while the second term represents the anomalous Hall effect~\cite{nagaosa2010,armitage2018a}.

In the following subsections, we provide a step-by-step derivation of Eq.~(\ref{eq:main_result}) and a closed-form expression of $\varepsilon_D(\omega)$. Then we discuss their physical consequences. We use SI units and $e^{-i\omega t}$ convention.

\subsection{A minimal low-energy model of Weyl semimetals}

\begin{figure}[htbp]
    \centering
    \includegraphics[width=0.7\textwidth]{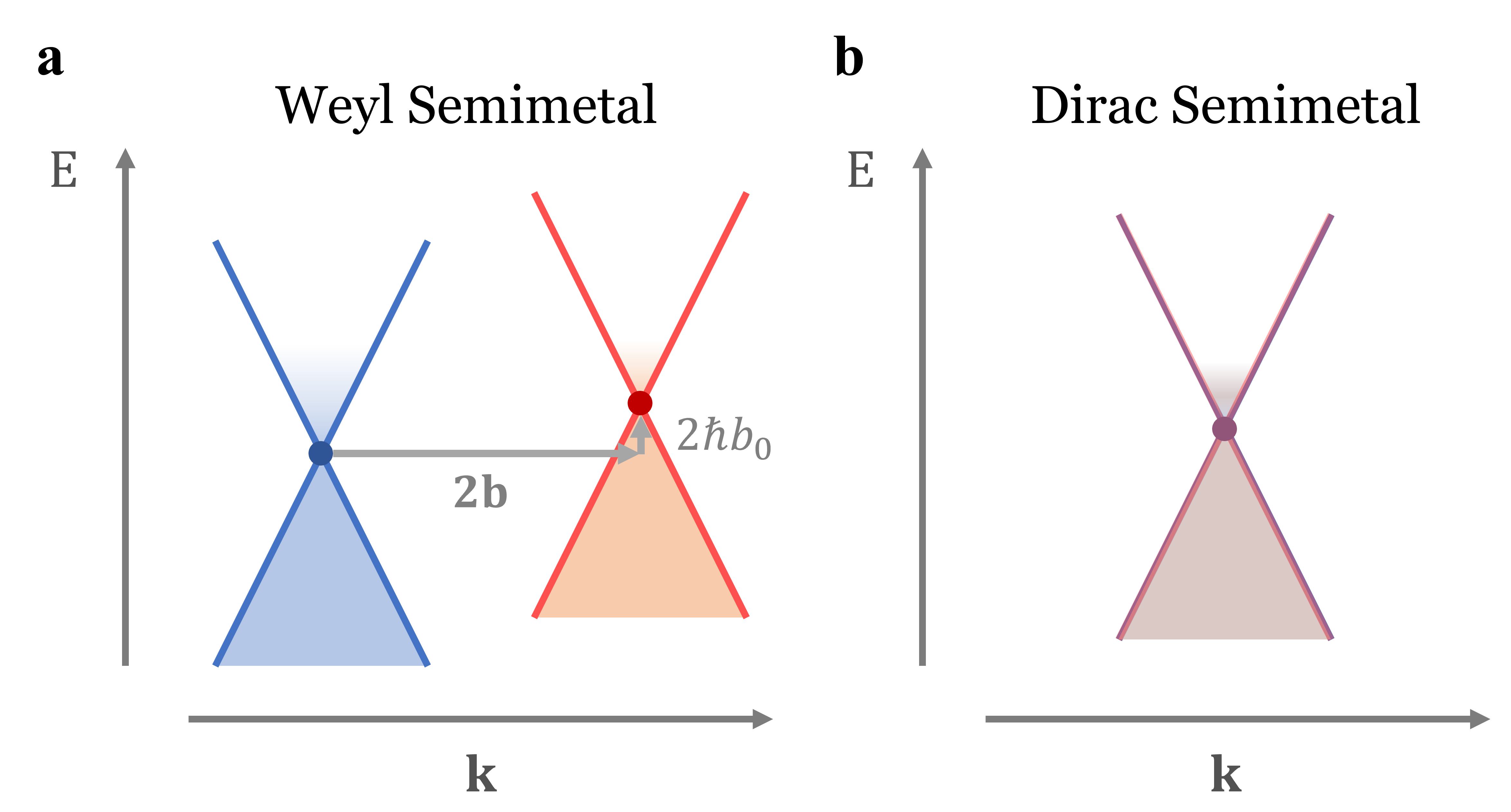}
    \caption{\csentence{A minimal low-energy model.} Schematic band dispersion for (a) a Weyl semimetal, (b) a Dirac semimetal. In (a), the blue and red lines correspond to right- and left-handed quasiparticles; the blue and red dots denote the right- and left-handed Weyl nodes. In (b), these lines/dots coalesce. }
    \label{fig:energy}
\end{figure}

According to the Nielsen-Ninomiya theorem~\cite{nielsen1981,nielsen1981a,nielsen1981b}, a Weyl semimetal always include an even number of Weyl nodes and the total chirality of all nodes must vanish. Hence, the simplest Weyl semimetal has a single pair of Weyl nodes. It can be described by the following minimal low-energy Hamiltonian:
\begin{equation}\label{eq:Hamiltonian_minimal}
    H = \begin{pmatrix}
    \hbar v_F \bm{\sigma} \cdot \left(\bm{k} + \bm{b}\right) - \hbar b_0 I & 0 \\
    0 & -\hbar v_F \bm{\sigma} \cdot \left(\bm{k} - \bm{b}\right) + \hbar b_0 I
    \end{pmatrix},
\end{equation}
which characterizes two Weyl nodes of opposite chirality that are separated by $2\bm{b}$ in wavevector and $2 \hbar b_0$ in energy. Here $\bm{b}$ is also known as the chiral shift~\cite{gorbar2009,gorbar2021}, $\bm{k}$ denotes the wavevector, $v_F$ denotes the Fermi velocity,  $\bm{\sigma}$ is the vector of Pauli matrices, and $I$ is the $2\times 2$ identity matrix.

From Eq.~(\ref{eq:Hamiltonian_minimal}), the band dispersion is determined as
\begin{equation}
    E_\lambda(\bm{k}) = - \lambda \hbar b_0 \pm \hbar v_F \left|\bm{k} + \lambda \bm{b}\right|,  
\end{equation}
where $\lambda = \pm 1$ is the node's chirality. Fig.~\ref{fig:energy} shows the scheme of band dispersion in the cases (a) $\bm{b}\neq \bm{0}$ and $b_0 \neq 0$ and (b) $\bm{b} = \bm{0}$ and $b_0 = 0$. The latter case, where the two Weyl nodes coalesce, is referred to as a Dirac semimetal.

We make a few remarks. First, the model is paradigmatic and highlights the two hallmark properties of Weyl semimetals: linear dispersion and chiral Weyl nodes. Second, the model suffices to describe the universal low-energy phenomena in  Weyl semimetals. It can be shown by explicit calculation that the general properties do not change if one uses a more realistic periodic two-band model~\cite{zyuzin2012,gorbar2021}. Third, nonzero $\bm{b}$ and $b_0$ require the breaking of time-reversal and parity-inversion symmetries, respectively. Fourth, we note that the opposite sign convention of $\bm{b}$ and $b_0$ has been used in some works~\cite{gorbar2021}. Finally, we note that the ideal magnetic Weyl semimetal with the minimum number of Weyl points may be realized in real materials such as K$_2$Mn$_3$(AsO$_4$)$_3$~\cite{nie2022}, or EuCd$_2$As$_2$ subjected to an external magnetic field~\cite{soh2019}.

\subsection{Axion electrodynamics}

To characterize the electromagnetic properties of the Weyl semimetal as described by Eq.~(\ref{eq:Hamiltonian_minimal}), we first briefly review the Lagrangian approach to Maxwell's equations~\cite{zangwill2013}. We start with the conventional Maxwell Lagrangian density:
\begin{equation}\label{eq:maxwell_lagrangian}
    \mathcal{L}_0(\bm{r},t) = \frac{\epsilon_0}{2} \bm{E}^2 - \frac{1}{2\mu_0} \bm{B}^2 -\rho \phi +\bm{J}\cdot \bm{A}.
\end{equation}
Here, the potentials $\phi(\bm{r},t)$ and $\bm{A}(\bm{r},t)$ are the independent variables; the electric field $\bm{E}(\bm{r},t)$ and the magnetic field $\bm{B}(\bm{r},t)$ are expressed in terms of potentials:
\begin{equation}\label{eq:fields_via_potentials}
    \bm{B} = \nabla \times \bm{A}, \qquad \bm{E} = -\nabla\phi - \frac{\partial{\bm{A}}}{\partial t}.  
\end{equation}
Equations~(\ref{eq:fields_via_potentials}) already imply   two of the four Maxwell's equations: 
\begin{equation}\label{eq:homogeneous_Maxwell_equations}
    \nabla \cdot \bm{B} = 0, \qquad \nabla \times \bm{E} + \frac{\partial \bm{B}}{\partial t}  = 0.
\end{equation} 
We define the action 
\begin{equation}
    S_0 = \int \mathrm{d} t \,\mathrm{d}^3 \bm{r} \,\mathcal{L}_0 (\bm{r},t),
\end{equation}
and require $S_0$ to be stationary with respect to the variations of $\bm{\phi}(\bm{r},t)$ and $\bm{A}(\bm{r},t)$. Then we obtain the other two of the four Maxwell's equations:  
\begin{equation}\label{eq:inhomogeneous_Maxwell_equations}
    \nabla \cdot \bm{E} = \frac{\rho}{\epsilon_0}, \qquad \nabla \times \bm{B} = \mu_0 \bm{J} + \frac{1}{c^2} \frac{\partial E}{\partial t}. 
\end{equation}
See Ref.~\cite[\S 24]{zangwill2013} for detailed derivation of Eq.~(\ref{eq:inhomogeneous_Maxwell_equations}).

Now, we introduce the formalism of axion electrodynamics~\cite{wilczek1987}. Axion electrodynamics was first proposed in high-energy physics to solve the strong CP problem in quantum chromodynamics~\cite{cheng1984}. In condensed matter physics, axion electrodynamics is used to understand the properties
of $^3$He~\cite{volovik2009}, topological insulators~\cite{qi2008a,qi2011,bernevig2013,hasan2015}, and Weyl semimetals~\cite{zyuzin2012,grushin2012,goswami2013,vazifeh2013}. For Weyl semimetals, axion electrodynamics describes the topological effects of Weyl node separation. The usage of axion electrodynamics in Weyl semimetals will be discussed in the next subsection.

Axion electrodynamics is generated by adding to Eq.~(\ref{eq:maxwell_lagrangian}) an additional term (the $\theta$ term)~\cite{wilczek1987,deng2021}:
\begin{equation}\label{eq:axion_lagrangian}
    \mathcal{L}_\theta = 2 \alpha \sqrt{\frac{\epsilon_0}{\mu_0}}\frac{\theta}{2\pi} \bm{E}\cdot \bm{B},
\end{equation}
where $\alpha = \frac{e^2}{4\pi\epsilon_0 \hbar c}$ is the fine structure constant, and $\theta (\bm{r},t)$ is a pseudoscalar field. Now
\begin{equation}
    S = \int \mathrm{d} t \,\mathrm{d}^3 \bm{r} \,\left[\mathcal{L}_0 (\bm{r},t) + \mathcal{L}_\theta (\bm{r},t)\right].
\end{equation}
This does not modify Eqs.~(\ref{eq:homogeneous_Maxwell_equations}), but changes Eqs.~(\ref{eq:inhomogeneous_Maxwell_equations}) into (see Ref.~\cite{armitage2019a} for detailed derivation):
\begin{align}
    \nabla \cdot \bm{E} &= \frac{\rho}{\epsilon_0} - 2c\alpha \nabla \left(\frac{\theta}{2\pi}\right) \cdot \bm{B}, \\
    \nabla \times \bm{B} &= \mu_0 \bm{J} + \frac{1}{c^2} \frac{\partial \bm{E}}{\partial t} + \frac{2\alpha}{c} \left[\frac{\partial}{\partial t}\left(\frac{\theta}{2\pi}\right) \bm{B} + \nabla \left(\frac{\theta}{2\pi}\right)\times \bm{E}\right].
\end{align}
These modified Maxwell's equations contain most (although not all) of the new physics of axion electrodynamics~\cite{tong2018}. Here we note that the $\theta$ term enters the equations only through derivatives. This is as expected since if $\mathcal{\theta}$ is constant~\cite{karch2009}, the $\theta$ term is a total derivative and irrelevant for the equation of motion.

There is a different yet equivalent way to describe axion electrodynamics~\cite{karch2009}, where one keeps the macroscopic Maxwell's equations in their original form:
\begin{align}
&\nabla \cdot \bm{B} = 0, \qquad \nabla \times \bm{E} = -\frac{\partial{\bm{B}}}{\partial t},  \\
 &\nabla \cdot \bm{D} = \rho, \qquad \nabla \times \bm{H} = \bm{J} + \frac{\partial \bm{D}}{\partial t}, 
\end{align}
with the modified constitutive relations:
\begin{equation}
    \bm{D} = \epsilon_0 \bm{E} + 2\alpha c \epsilon_0 \frac{\theta}{2\pi} \bm{B}, \qquad \bm{H} = \frac{1}{\mu_0} \bm{B} - 2\alpha \frac{1}{c \mu_0} \frac{\theta}{2\pi} \bm{E}. 
\end{equation}

There is yet another way to describe axion electrodynamics if $\theta(\bm{r},t)$ satisfies the additional condition:
\begin{equation}\label{eq:additional_condition}
\nabla \left[\frac{\partial \theta}{\partial t} (\bm{r},t)\right] = \bm{0}.
\end{equation}
Then one can express  Maxwell's macroscopic equations in the frequency domain: 
\begin{align}
&\nabla \cdot \bm{B} = 0, \qquad \nabla \times \bm{E} = i\omega\bm{B},  \\
 &\nabla \cdot \bm{D} = \rho, \qquad \nabla \times \bm{H} = \bm{J} - i\omega \bm{D}.
\end{align}
with the constitutive relation~\cite{hofmann_surface_2016}:
\begin{equation}\label{eq:constitutitve_axion}
    \bm{D} = \varepsilon_0 \bm{E} + 2 \alpha c \varepsilon_0 \frac{1}{-i\omega}\left[\frac{\partial}{\partial t} \left(\frac{\theta}{2\pi}\right)\bm{B}+\nabla(\frac{\theta}{2\pi})\times \bm{E}\right],  \quad   \bm{H} = \frac{1}{\mu_0}\bm{B}.
\end{equation}
Such a description is equivalent to the previous two under the assumption of Eq.~(\ref{eq:additional_condition}). It will be useful later when we discuss Weyl semimetals.

\subsection{The axion term}
\label{subsec:axion_weyl}

It is known that the effects of Weyl node separation are fully described by an axion term
\begin{equation}\label{eq:theta_field}
\theta(r,t) = 2 \bm{b}\cdot \bm{r} - 2 b_0 t.   
\end{equation}
Substituting Eq.~(\ref{eq:theta_field}) into Eq.~(\ref{eq:constitutitve_axion}) and replacing $\varepsilon_0$ with $\varepsilon_D(\omega)$, we obtain Eq.~(\ref{eq:main_result}). (Note that Eq.~(\ref{eq:theta_field}) satisfies the assumption of Eq.~(\ref{eq:additional_condition}).)

The above result is standard; its derivation is beyond the scope of this tutorial. We only note that the derivation uses Fujikawa's method in the fermion path integral formulation; the $\theta$ term corresponds to the chiral anomaly, a type of quantum anomalies that arises from a nontrivial Jacobian in
the change of path integral variables.
We refer readers to Ref.~\cite{zyuzin2012} for the original derivation of Eq.~(\ref{eq:theta_field}), Ref.~\cite{witten2016a} for fermion path integrals, and Ref.~\cite{fujikawa2013} for quantum anomalies and Fujikawa's method. 

\subsection{The Dirac term}

Now we derive the Dirac term $\varepsilon_D(\omega)$ in Eq.~(\ref{eq:main_result}), i.e., the permittivity of a Dirac semimetal (Fig.~\ref{fig:energy}b). The Dirac semimetal can be described by a simple two-band Hamiltonian:
\begin{equation}\label{eq:Hamiltonian_Dirac}
    H_D = \hbar v_F \bm{k}\cdot \bm{\sigma},
\end{equation}
while the spin degeneracy is taken into account by a degeneracy factor $g=2$. 

From the Hamiltonian Eq.~(\ref{eq:Hamiltonian_Dirac}), we obtain the low-energy spectra:
\begin{equation}
    E_{\bm{k},s} = s \hbar v_F |\bm{k}|, 
\end{equation}
where $s=\pm 1$ denote the band indices. The eigenstates are:
\begin{equation}
    \ket{\bm{k},+} = \begin{pmatrix} 
    \cos{\frac{\vartheta}{2}} \\
    e^{i\varphi} \sin{\frac{\vartheta}{2}}
    \end{pmatrix} , \qquad  
    \ket{\bm{k},-} = \begin{pmatrix}  
    -\sin{\frac{\vartheta}{2}} \\
    e^{i\varphi} \cos{\frac{\vartheta}{2}}
    \end{pmatrix},     
\end{equation}
where $\vartheta$ and $\varphi$ are respectively the polar and azimuthal angles of the three-dimensional $\bm{k}$.

Then $\varepsilon_D(\omega)$ can be determined by the standard linear response theory~\cite{kubo1957,greenwood1958}:
\begin{equation}\label{eq:eps_dirac}
\varepsilon_D(\omega) = \varepsilon_b + i \frac{\sigma(\omega)}{\omega},
\end{equation}
where $\varepsilon_b$ is the background permittivity due to the other bands and $\sigma(\omega)$ is the dynamic conductivity tensor due to the Dirac cone. Assuming noninteracting electrons and local response, $\sigma(\omega)$ is given by Kubo-Greenwood formula~\cite{moseley1978,kotov_giant_2018}:
\begin{equation}
    \sigma_{\alpha\beta}(\omega) = \frac{-ie^2 g\hbar}{V} \sum_{\bm{k},s,s'} \frac{n(E_{\bm{k},s})-n(E_{\bm{k},s'})}{E_{\bm{k},s}-E_{\bm{k},s'}} \frac{\bra{\bm{k} s}\hat{v}_\alpha\ket{\bm{k} s'}\bra{\bm{k} s'}\hat{v}_\beta\ket{\bm{k} s}}{\hbar(\omega+i0)+E_{\bm{k},s}-E_{\bm{k},s'}}.
\end{equation}
Here $\alpha, \beta = x,y,z$, $\omega$ is the incident light frequency, $V$ is the volume, $\hat{v}_\alpha = \frac{1}{\hbar} \frac{\partial H_D}{\partial k_\alpha} = v_F \sigma_\alpha$ is the velocity operator, $\ket{\bm{k}s}$ and $\ket{\bm{k}s'}$ are the initial and final electronic state, and 
\begin{equation}
n(E) = \frac{1}{e^{(E-E_F)/k_B T}+1}    
\end{equation} 
is the Fermi distribution where $E_F$ is the Fermi energy and $T$ is the temperature.  

Since the Hamiltonian Eq.~(\ref{eq:Hamiltonian_Dirac}) is isotropic in $\bm{k}$, the conductivity tensor $\sigma(\omega)$ is also isotropic and can be treated as a scalar. Detailed calculation shows~\cite{kotov_giant_2018}:
\begin{equation}
\label{eq:sigma}
    \sigma(\omega) = \frac{e^2}{\hbar}\frac{g k_F}{24\pi}\Omega \tilde{G}(\Omega/2) + i\frac{e^2}{\hbar}\frac{g k_F}{24\pi^2}\left\{\frac{4}{\Omega}\left[1+\frac{\pi^2}{3}(\frac{k_B T}{E_F})^2\right]+8\Omega\int_0^{\varepsilon_c} \frac{\tilde{G}(\varepsilon)-\tilde{G}(\Omega/2)}{\Omega^2-4\varepsilon^2}\varepsilon\mathrm{d}\varepsilon\right\},
\end{equation}
where $k_F = E_F/\hbar v_F$, $\Omega = \hbar (\omega + i \tau^{-1}) / E_F$,  $\tau^{-1}$ is the Drude damping rate, $\varepsilon_c = E_c/E_F$, and $E_c$ is the cutoff energy beyond which the band dispersion is no longer linear.  Moreover, the function $\tilde{G}(x) $ in Eq.~(\ref{eq:sigma}) reads:
\begin{equation}
\tilde{G}(x) \equiv n(-xE_F) - n(xE_F) = \frac{\sinh{(xE_F/k_B T)}}{\cosh{(E_F/k_B T)}+\cosh{(xE_F/k_B T)}}.
\end{equation}

The physical meaning of Eq.~(\ref{eq:sigma}) is more transparent in the low-temperature limit when $k_B T\ll E_F$. Then  $\tilde{G}(\Omega/2)\rightarrow \Theta(\Omega-2)$ where $\Theta(\cdot)$ is the Heaviside step function, and thus:
\begin{align}
    \sigma(\omega) = \frac{e^2}{\hbar}\frac{g k_F}{24\pi}\Omega\, \Theta(\Omega-2) + i \frac{e^2}{\hbar}\frac{g k_F}{24\pi^2}\left[\frac{4}{\Omega}-\Omega \ln{\frac{4\varepsilon_c^2}{|\Omega^2 - 4|}}\right] 
\end{align}
For the real part of  $\sigma(\omega)$, the step function captures the interband absorption when $E>2E_F$. For its imaginary part, the first term  is the Drude term due to the intraband transition, while the second term is the correction due to the interband transitions.

\subsection{Giant optical nonreciprocity}\label{subsec:optical_nonreciprocity}

Now we can discuss one unique property of magnetic Weyl semimetals: giant optical nonreciprocity without an external magnetic field. 
Reciprocity is a fundamental internal symmetry of Maxwell's equations~\cite{lorentz_theorem_1896, potton_reciprocity_2004,caloz_electromagnetic_2018,asadchy_tutorial_2020,guo2022a}. It imposes direct constraints on basic optical phenomena including transmission~\cite{haus1984}, reflection~\cite{guo2022c}, absorption, and emission~\cite{kirchhoff1860,greffet2018}. Conversely, breaking reciprocity enables significant new opportunities in photonic applications such as isolation~\cite{yu_complete_2009,jalas_what_2013}, circulation~\cite{wang2005},
robust topological transport~\cite{wang2009b,fang_realizing_2012}, and violation of Kirchhoff’s law of thermal radiation~\cite{Zhu2014a,guo2022b}. 

Optical materials with significant nonreciprocal responses are rare. The most used ones are magneto-optical materials~\cite{zvezdin_modern_1997}. Under an external magnetic field, this type of material has an asymmetric dielectric tensor $\varepsilon \neq \varepsilon^T$ that breaks reciprocity ($\varepsilon^T$ denotes the transpose of $\varepsilon$). The strength of nonreciprocal effects depends on the degree of asymmetry of $\varepsilon$~\cite{zvezdin_modern_1997}:
\begin{equation}
\gamma = \frac{\left\|\varepsilon-\varepsilon^T\right\|}{\left\|\varepsilon+\varepsilon^T\right\|} ,
\end{equation}
where $\|\cdot\|$ denotes the matrix norm. For magneto-optical materials, 
\begin{equation}
\gamma \sim \frac{\omega_c}{\omega}\,,
\end{equation}
where $\omega$ is the light frequency, $\omega_c = eB/m^*$ is the cyclotron frequency, $m^*$ is the effective electron mass, $e$ is the electron charge, and $B$ is the external magnetic field~\cite{zvezdin_modern_1997,Zhu2014a}. For the typical magnetic field $B\sim \SI{1}{T}$, $\omega_c\sim \SI{1}{THz}$, thus $\gamma\sim 0.001$-$0.01$ at optical frequencies. Therefore, the nonreciprocal effect is weak in magneto-optical materials.

\begin{figure}[tb]
    \centering
    \includegraphics[width=0.95\textwidth]{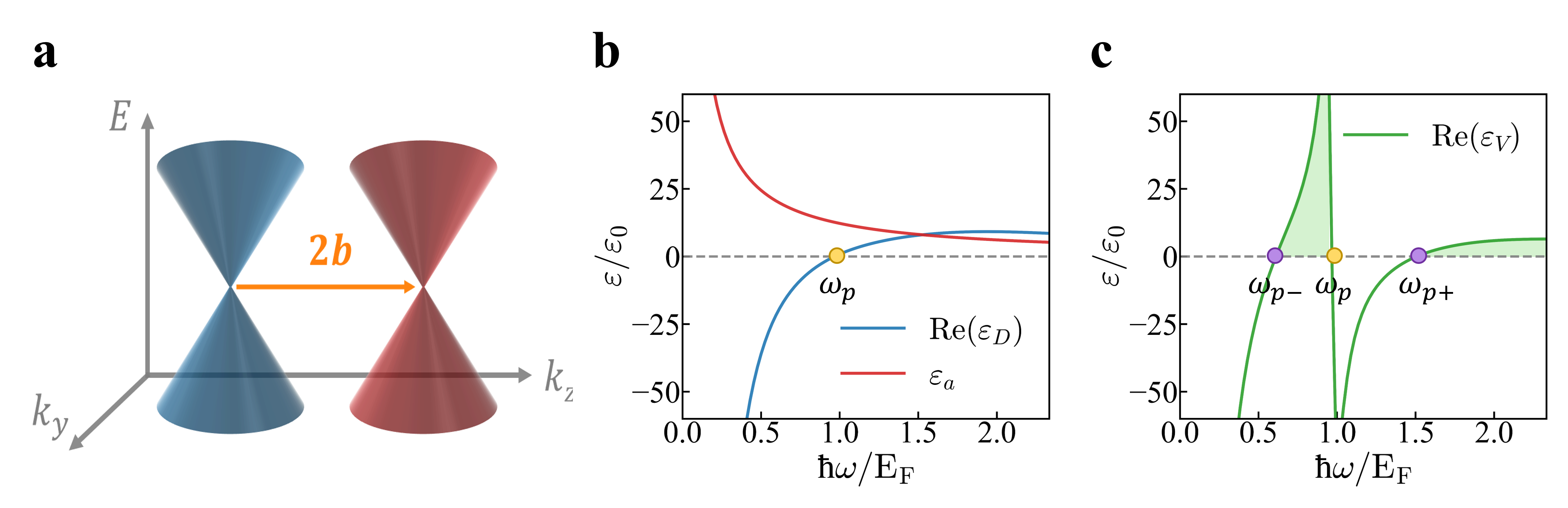}
    \caption{\csentence{Permittivity of a magnetic Weyl semimetal.} (a) Band structure of a Weyl semimetal with a chiral shift $2\bm{b}$. (b) $\varepsilon_a$ and the real part of $\varepsilon_D$ in the permittivity tensor of Eq.~(\ref{eq:dielectric_tensor}). $\varepsilon_0$ denotes the vacuum permittivity. (c) The real part of the Voigt permittivity $\varepsilon_V$ in Eq.~(\ref{eq:Voigt_permittivity}). The regions where $\operatorname{Re}(\varepsilon_V)>0$ are shaded by color. Figures (a, b) are reproduced with permission from Ref.~\cite{zhao_axion-field-enabled_2020}. Copyright 2020 American Chemical Society. }
    \label{fig:scheme}
\end{figure}

In contrast, magnetic Weyl semimetals can exhibit extremely large nonreciprocal responses, with $\gamma\sim1$ at optical frequencies. As an illustration, we consider a Weyl semimetal with $\bm{b}$ along the $z$-direction and $b_0 = 0$ (Fig.~\ref{fig:scheme}a). It has a permittivity tensor:
\begin{equation}
\label{eq:dielectric_tensor}
    \varepsilon(\omega) = 
    \begin{pmatrix}
    \varepsilon_D & i \varepsilon_a & 0 \\
    -i\varepsilon_a & \varepsilon_D & 0 \\
    0 & 0 & \varepsilon_D 
    \end{pmatrix}, 
\end{equation}
where $\varepsilon_D(\omega)$ is given by Eqs.~(\ref{eq:eps_dirac}) and~(\ref{eq:sigma}), and
\begin{equation}
    \varepsilon_a(\omega) = \frac{be^2}{2\pi^2\hbar\omega}.
\end{equation}
The permittivity tensor $\varepsilon(\omega)$ has the typical form of a gyrotropic medium~\cite{zvezdin_modern_1997}. Following Ref.~\cite{kotov_giant_2018}, we use the parameters $\varepsilon_b/\varepsilon_0 = 6.2$, $\xi_c = 3$, $g=2$, $\tau= \SI{1000}{fs}$, $b = \SI{2e9}{\per m}$, $v_F = \SI{0.83e5}{m/s}$, $T=\SI{300}{K}$, and $E_F = \SI{0.15}{eV}$. Fig.~\ref{fig:scheme}b plots the calculated $\varepsilon_D$ and $\varepsilon_a$. The magnitude of $\varepsilon_a$ is comparable to $\varepsilon_D$ over a broad frequency range. Hence $\gamma \sim |\varepsilon_a / \varepsilon_D| \sim 1$. Such a strong nonreciprocity originates from the anomalous Hall effect induced by the Weyl node separation~\cite{armitage2018a,gorbar2021,han2022}. This mechanism is fundamentally different from the cyclotron mechanism for magneto-optical materials. In particular, it requires no external magnetic field.

\subsection{Electromagnetic waves in the bulk medium}\label{subsec:EM_wave_in_bulk}

The strong gyrotropy significantly affects the behavior of light in magnetic Weyl semimetals.
As the first example, we study the electromagnetic waves in the bulk medium as described by $\varepsilon(\omega)$ in Eq.~(\ref{eq:dielectric_tensor}). The results will be useful later in discussing many applications. 

We consider the Voigt configuration, i.e.,~the wavevector $\bm{k}$ is in the plane perpendicular to the axis of gyration $\bm{b}$. Such a configuration is mirror symmetric with respect to that plane, thus waves can be decoupled into TE and TM modes~\cite{huang2008}. TE/TM modes have the electric/magnetic fields parallel to $\bm{b}$; their dispersion relations can be expressed as:
\begin{align}
\text{TE:} \qquad k^2 = \omega^2 \mu_0 \varepsilon_D(\omega), \label{eq:dispersion_TE} \\
\text{TM:} \qquad k^2 = \omega^2 \mu_0 \varepsilon_V(\omega). \label{eq:dispersion_TM}
\end{align}
where we have introduced the Voigt permittivity for TM modes as:
\begin{equation}\label{eq:Voigt_permittivity}
\varepsilon_V(\omega) = \varepsilon_D - \frac{\varepsilon_a^2}{\varepsilon_D}. 
\end{equation}
Fig.~\ref{fig:scheme}c plots the calculated $\varepsilon_V(\omega)$ with the parameters as described in Sec.~\ref{subsec:optical_nonreciprocity}.

We note that the gyrotropic medium behaves differently for the two modes due to the different permittivity components: $\varepsilon_D$ for TE modes in Eq.~(\ref{eq:dispersion_TE}) and $\varepsilon_V$ for TM modes in Eq.~(\ref{eq:dispersion_TM}). Moreover, the two permittivities have different frequency dependence. For TE modes, $\varepsilon_D$ is not affected by the gyrotropy (Fig.~\ref{fig:scheme}b). It monotonically increases with the frequency and has a plasma frequency $\omega_p$. For TM modes, $\varepsilon_V$ is affected by the gyrotropy (Fig.~\ref{fig:scheme}c). The existence of the chiral shift $2\bm{b}$ splits $\varepsilon_V$ into two branches which are separated by $\omega_p$. In each branch, $\varepsilon_V$ monotonically increases with the frequency and has an effective plasma frequency $\omega_{p-}$ and $\omega_{p+}$, respectively. Consequently, when $\omega_{p-}<\omega< \omega_p$, $\operatorname{Re}(\varepsilon_D)$ is negative while $\operatorname{Re}(\varepsilon_V)$ is positive;  when $\omega_{p}<\omega< \omega_{p+}$, $\operatorname{Re}(\varepsilon_D)$ is positive while $\operatorname{Re}(\varepsilon_V)$ is negative.

\subsection{Nonreciprocal surface plasmon polaritons}\label{subsec:plasmons}

As the second example, we study the surface plasmon polaritons of a Weyl semimetal~\cite{hofmann_surface_2016}. The results will also be useful for later applications. 

\begin{figure}[htbp]
    \centering
    \includegraphics[width=0.95\textwidth]{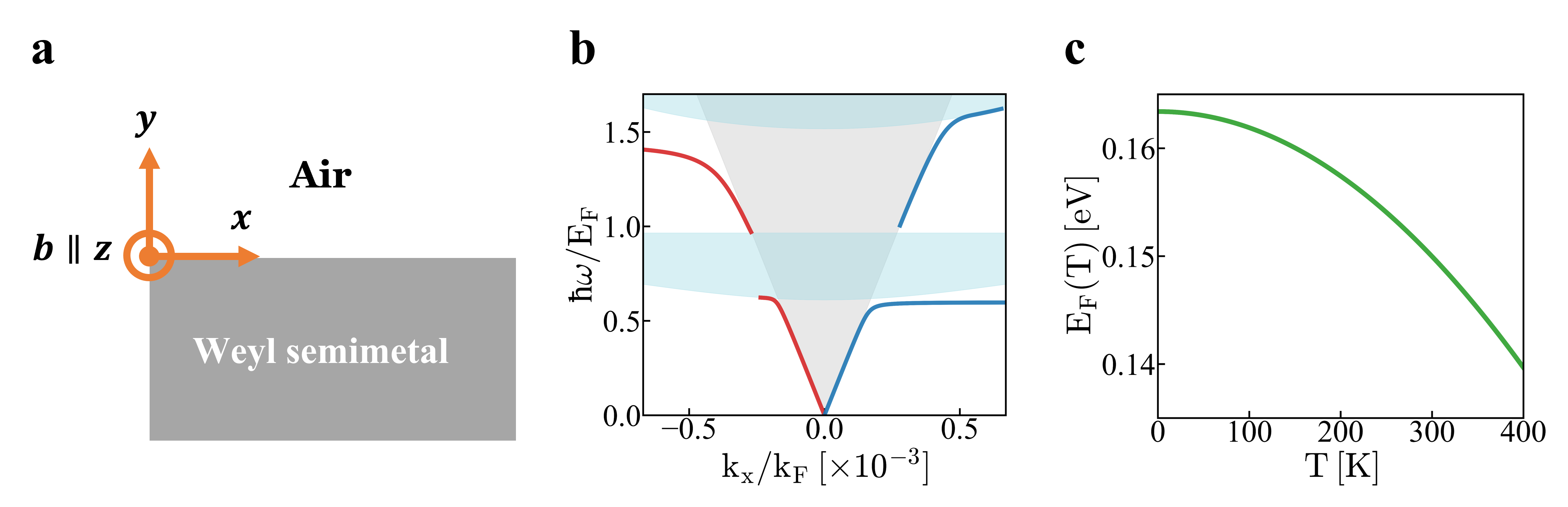}
    \caption{\csentence{Nonreciprocal surface plasmon polaritons.} (a) An interface between air and a magnetic semimetal. (b) Dispersion relation of the surface plasmon polaritons. The gray region shows the light cone of the vacuum. The blue region shows the continuum of the propagating modes in the bulk. $k_F = E_F/\hbar v_F$ is the Fermi wavevector. (c) The temperature dependence of the chemical potential $E_F(T)$.  Figures (b, c) are reproduced with permission from Ref.~\cite{zhao_axion-field-enabled_2020}.  Copyright 2020 American Chemical Society.} 
    \label{fig:spp}
\end{figure}

We consider a planar interface between air and a  semi-infinite magnetic Weyl semimetal with a chiral shift $2\bm{b}$ along the $z$ direction parallel to the interface (Fig.~\ref{fig:spp}a). Such an interface supports nonreciprocal surface plasmon polaritons, whose band dispersion along the $x$ direction is determined by:~\cite{kotov_giant_2018,abdol_surface_2019}
\begin{equation}\label{eq:dispersion}
k _ { y } +\varepsilon _ {V}  k _ {y0 }  - k _ { x }\dfrac{\varepsilon _ { a }} {\varepsilon _ {D}}= 0\,,
\end{equation}
where $k_0(\omega) = \omega/c$ is the wavenumber in vacuum, $k_x$ is the $x$-component of the wavevector,  $k_{y0} = \sqrt{k_x^2- k_0^2}$ and $k_y= \sqrt{k_x^2-\epsilon_V k_0^2}$ are the $y$-component of the wavevector in air and in the Weyl semimetal, respectively. The last term in Eq.~(\ref{eq:dispersion}) shows that when the sign of $k_x$ flips, the frequency of the surface plasmon polaritons will be different, i.e.,~$\omega(k_x)\neq \omega(-k_x)$. The asymmetry of the dispersion is induced by the gyrotropy; it occurs only when $\varepsilon_a \neq 0$.

Fig.~\ref{fig:spp}b shows the calculated band dispersion of the surface plasmon polaritons, together with the continuum region of bulk modes (in light blue) and the light cone of vacuum (in light gray). We note that the surface plasmon polaritons approximately occupy the frequency ranges $\omega<\omega_{p-}$ and $\omega_p < \omega<\omega_{p+}$; the bulk modes occupy the frequency ranges $\omega_{p-}<\omega<\omega_{p}$ and $\omega>\omega_{p+}$. This is as expected from the frequency dependence of $\varepsilon_V$ in Fig.~\ref{fig:scheme}c. The dispersion of the surface plasmon polaritons is clearly asymmetric near the effective plasma frequencies $\omega_{p-}$ and $\omega_{p+}$. Away from the effective plasma frequencies, the dispersion becomes approximately symmetric. The frequency range where the dispersion is evidently asymmetric measures the strength of nonreciprocity. The plots confirm the strong nonreciprocity of a magnetic Weyl semimetal without an external magnetic field. As a comparison, to obtain a similar strength of nonreciprocity, conventional magneto-optical materials, such as InSb, need an unrealistically
large external magnetic field of about \SI{100}{T}~\cite{zhao_axion-field-enabled_2020,asadchy_sub-wavelength_2020}. This is consistent with our order-of-magnitude comparison of $\gamma$ in Sec.~\ref{subsec:optical_nonreciprocity}. 

\subsection{Tunable Fermi level}\label{subsec:tunable_Fermi_level}

Another unique property of Weyl semimetals is the large tunability of the Fermi level $E_F$ (also referred to as the chemical potential). The variation of the Fermi level can significantly affect the optical conductivity according to Eq.~(\ref{eq:sigma}). This leads to greatly tunable optical phenomena. The Fermi level of a Weyl semimetal can be adjusted in different ways. Below we discuss two approaches. 

First, the Fermi level can be tuned by electric gating. This is because an undoped Weyl semimetal features a vanishing density of states at the energy of Weyl points. Therefore, the carrier concentrations are much lower than metals and thus are more easily depleted. 
Second, the Fermi level can be tuned by thermal tuning. This is because a Weyl semimetal has a small and nonconstant density of states and a linear dispersion. Consequently, its Fermi level  is strongly temperature dependent. The Fermi level as a function of temperature can be calculated from the requirement of charge conservation~\cite{Ashby2014}:
\begin{equation}
E_F ( T ) = \frac { 2 ^ { 1 / 3 } \left( 9 E_F ( 0) ^ { 3 } + \sqrt { 81 E_F(0) ^ { 6 } + 12 \pi ^ { 6 } k_B^6 T ^ { 6 } } \right) ^ { 2 / 3 } - 2 \pi ^ { 2 } 3 ^ { 1 / 3 } k_B^2 T^2 } { 6 ^ { 2 / 3 } \left( 9 E_F(0)^ { 3 } + \sqrt { 81 E_F(0)^ { 6 } + 12 \pi ^ { 6 } k_B^6 T ^ { 6 } } \right) ^ { 1 / 3 } }.
\end{equation}
Fig.~\ref{fig:spp}c shows $E_F(T)$, where we set $E_F (0)= 0.163 $~eV such that $E_F (300$K$)= 0.150$~eV. Note that $E_F(T)$ decreases as $T$ increases. Consequently, the dielectric tensor and optical properties of the Weyl semimetal are also temperature dependent.

\subsection{Effects of Fermi arcs}\label{subsec:Fermi-arc}

As we mentioned in Sec.~\ref{subsec:Weyl_semimetals},  Weyl semimetals feature not only Weyl nodes in the bulk band structure but also Fermi arc surface states. The existence of Fermi arc surface states can modify the boundary conditions of Maxwell's equations, which can affect the optical properties of Weyl semimetals. Depending on the problems, the modification may be negligible in some cases but may be significant in other cases.

So far, we have focused on the optical effects that arise from the nontrivial \textit{bulk} electronic states in Weyl semimetals. Now we point to some works that highlight the optical effects of Fermi arcs. Ref.~\cite{huang2022a} provides a semiclassical approach to understanding surface Fermi arcs in Weyl semimetals. Ref.~\cite{chen_optical_2019} provides a detailed calculation of optical properties of Weyl semimetals including the effects of both bulk states and Fermi arc surface states, and compares their contributions to bulk and surface conductivity tensors. Refs.~\cite{song2017,losic_coupling_2019} study ``Fermi arc plasmons" that originate from the  hybridization of collective modes associated with Fermi arc carriers and bulk carriers. Such an unusual surface plasmon mode exhibits a hyperbolic band dispersion~\cite{gomez-diaz_hyperbolic_2015}. Ref.~\cite{zhang2021d} systematically studies both the bulk plasmons and the Fermi-arc plasmons over opposite surfaces of a Weyl semimetal slab.

\section{Photonic applications and devices}\label{sec:dev}

In this section, we survey various photonic applications of Weyl semimetals. These applications utilize nontrivial linear and nonlinear optical effects of Weyl semimetals. For each application, we discuss both the physical mechanism and the device configuration.

\subsection{Linear optical  effects}\label{subsec:linear_effects}
\subsubsection{Nonreciprocal optical components}\label{subsec:nonreciprocal_components}

Nonreciprocal optical components, such as isolators~\cite{jalas_what_2013}, circulators~\cite{kurokawa1969}, and nonreciprocal waveguides, are crucial in many photonic applications including optical circuits~\cite{dotsch_applications_2005,shoji_magneto-optical_2014}, and lasers~\cite{kravtsov_nonreciprocal_1999}. As we discussed in Sec.~\ref{subsec:optical_nonreciprocity}, magnetic Weyl semimetals can exhibit giant optical nonreciprocity without an external magnetic field~\cite{hofmann_surface_2016,kotov_giant_2018} thanks to the anomalous Hall effect~\cite{nagaosa2010,liu2018,soh2019,belopolski2019,morali2019,liu2019}. Hence, magnetic Weyl semimetals can be used to construct efficient and  compact nonreciprocal optical components.  Here, we review three examples of nonreciprocal optical components based on magnetic Weyl semimetals.

\begin{figure}[htbp]
    \centering
    \includegraphics[width=0.95\textwidth]{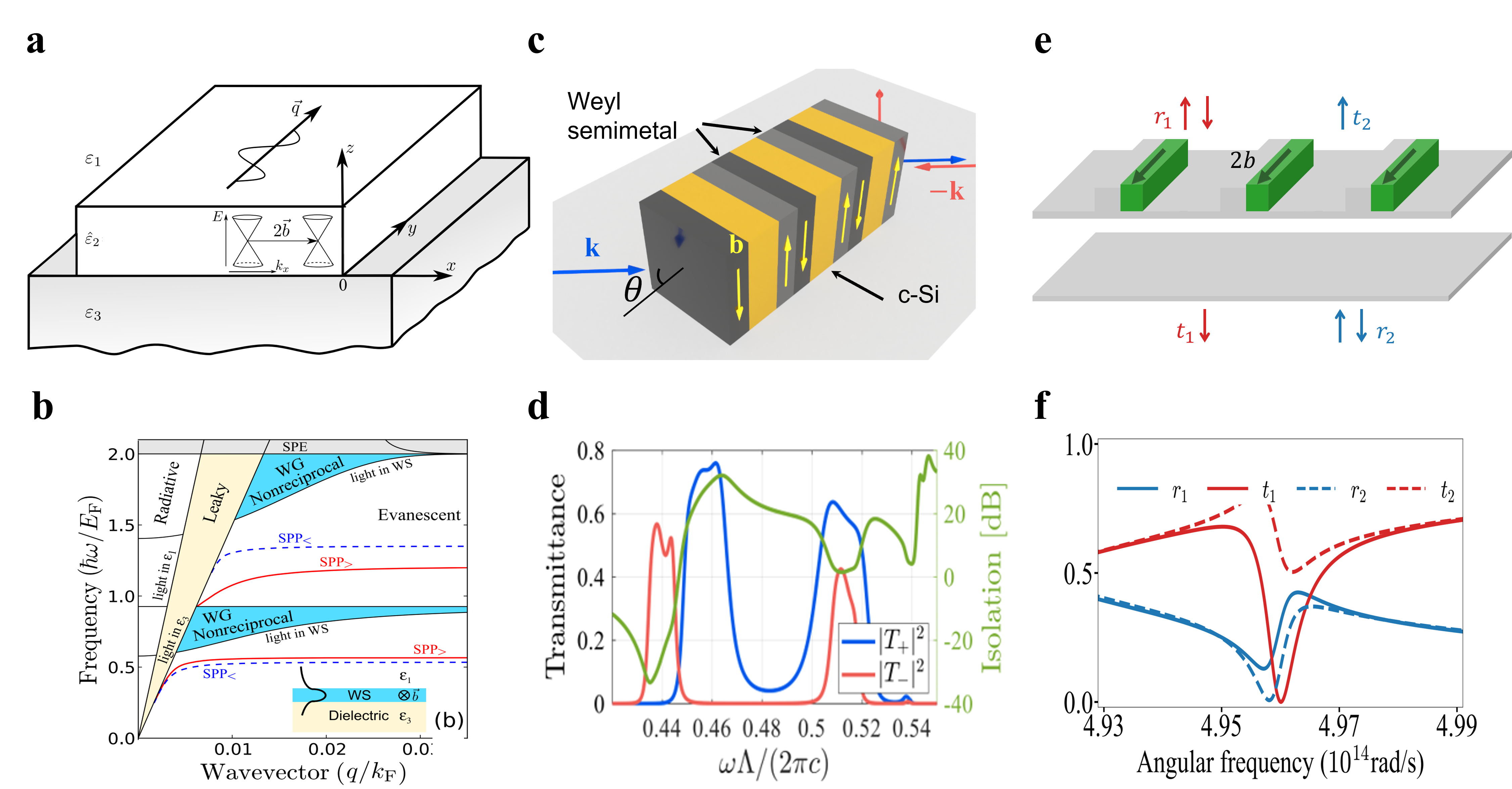}
    \caption{\csentence{Nonreciprocal optical components.} (a, b) A nonreciprocal waveguide based on magnetic Weyl semimetals. (a) The geometry consists of a Weyl semimetal film surrounded by two semi-infinite dielectrics. (b) The light dispersion at the waveguide. The red lines (${\rm SPP}_>$) and blue lines (${\rm SPP}_<$) denote the forward and backward nonreciprocal surface plasmon polaritons, respectively. The cyan regions denote the nonreciprocal waveguide modes. (c, d) A Voigt isolator based on magnetic Weyl semimetals. (c)  The geometry consists of three unit cells. Each unit cell comprises three material layers. The dark and bright grey layers are made of Weyl semimetals with opposite chiral shifts as indicated by the yellow arrows. The yellow layer is made of crystalline silicon (c-Si). The blue and red arrows denote forward and backward illuminations, respectively, at an incident angle $\theta = 45.6^\circ$. (d) The transmittance spectra in the forward (blue) and backward (red) directions. Also shown is the isolation ratio (green).  (e, f) An isolator based on a photonic crystal slab made of  Weyl semimetal and silicon. 
    (e) The geometry consists of a photonic crystal slab made of Weyl semimetals (green) and silicon (gray) and a uniform silicon slab separated by an air gap. (f) The reflectance and transmittance spectra in the forward (red) and backward (blue) directions. Figures are reproduced with permission from (a, b) Ref.~\cite{kotov_giant_2018}, Copyright 2018 American Physical Society; (c, d) Ref.~\cite{asadchy_sub-wavelength_2020}, Copyright 2020 John Wiley and Sons; (e, f) Ref.~\cite{park_violating_2021}, Copyright 2021 American Chemical Society. 
}
    \label{fig:isolators}
\end{figure}

In Ref.~\cite{kotov_giant_2018}, Kotov et al.~design a nonreciprocal waveguide using magnetic Weyl semimetals. As shown in Fig.~\ref{fig:isolators}a, the structure consists of a Weyl semimetal thin film sandwiched by two dielectrics. The chiral shift $2\bm{b}$ is along the $x$ direction. Fig.~\ref{fig:isolators}b shows the dispersion diagram of the TM-polarized light propagating in the $y$ direction. The structure supports two different types of guided modes depending on the light frequency $\omega$: When $\omega<\omega_{p-}$ or $\omega_p < \omega<\omega_{p+}$, the structure supports nonreciprocal surface plasmon polaritons; when $\omega_{p-}<\omega<\omega_{p}$ or $\omega > \omega_{p+}$, the structure supports nonreciprocal waveguide modes. Such behavior is similar to that shown in Fig.~\ref{fig:spp}b and is as expected from the frequency dependency of $\varepsilon_V$ in Fig.~\ref{fig:scheme}c. The dispersions of these guided modes can be tuned via gating between the THz and mid-IR ranges.

In Ref.~\cite{asadchy_sub-wavelength_2020}, Asadchy et al.~design optical isolators in the mid-IR range using magnetic Weyl semimetals, with both Faraday and Voigt geometries. The Faraday isolator consists of a Weyl semimetal slab sandwiched by two twisted polarization filters and operates at the normal incidence. The chiral shift $2 \bm{b}$ is along the direction perpendicular to the slab. The Voigt isolator (Fig.~\ref{fig:isolators}c) is a finite one-dimensional photonic crystal with a three-layer unit cell consisting of Weyl semimetal, dielectric, and Weyl semimetal; the chiral shifts of the two Weyl semimetal layers are both parallel to the slab but in the opposite directions. Such a structure is designed for optical isolation at oblique incidence. It breaks all the symmetries that preclude isolation~\cite{figotin_nonreciprocal_2001}. Numerical results show  that three unit cells suffice to achieve high isolation and low insertion loss. Remarkably, the total thickness of the device is less than $1.4$ times the  operation wavelengths, and is three orders of magnitude smaller than that of conventional isolators based on magnetooptical materials. Fig.~\ref{fig:isolators}d shows the calculated power transmittance spectra in the forward and backward directions at the incident angle of $\theta = \SI{45.6}{\degree}$. The two spectra are significantly different. Near the resonance frequency, the device achieves greater than $\SI{30}{dB}$ isolation and less than $\SI{1.2}{dB}$ insertion loss. In Ref.~\cite{li_strong_2021}, Li et al.~demonstrate a similar isolation performance using a one-dimensional photonic crystal containing a defect layer made of magnetic Weyl semimetals.

In Ref.~\cite{park_violating_2021}, Park et al.~design a photonic crystal slab structure made of magnetic Weyl semimetal and silicon, which can achieve optical isolation at normal incidence without the need for polarizers. As shown in Fig.~\ref{fig:isolators}e, the structure consists of a photonic crystal slab and a uniform dielectric slab separated by an air gap. The total thickness of the device is less than the operational wavelength. Such a structure hosts guided resonances~\cite{Fan2002} that can greatly enhance the nonreciprocal effects. Fig.~\ref{fig:isolators}f shows the calculated power transmission and reflection spectra for the TM polarized light in the forward and backward directions. The spectra exhibit high contrast near the guided resonance.

\subsubsection{Polarization filters}\label{polarization_filters}

\begin{figure}[htbp]
    \centering
    \includegraphics[width=0.8\textwidth]{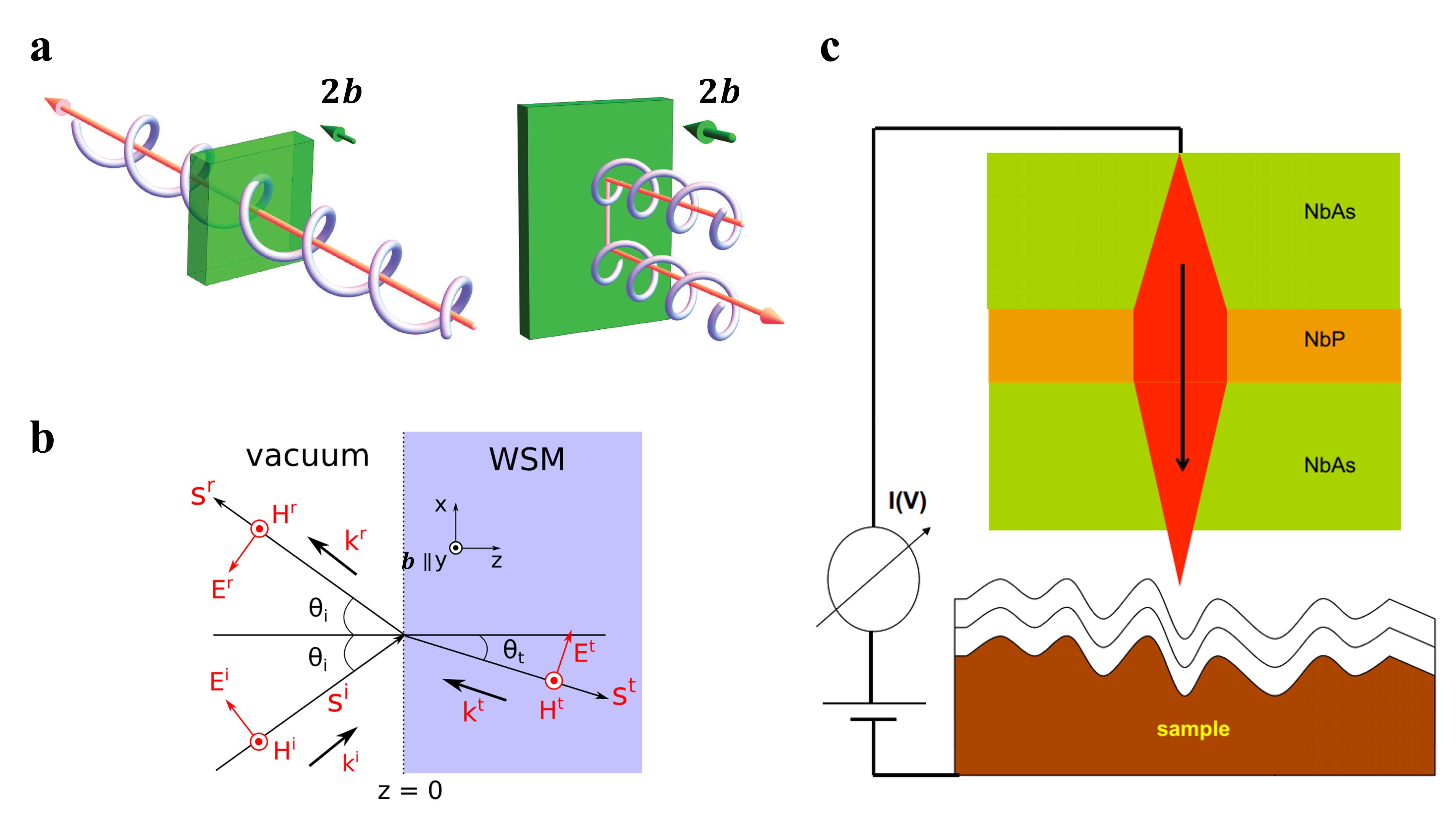}
    \caption{\csentence{Polarization filters and negative refraction.} (a) A circular polarization filter made of a magnetic Weyl semimetal slab in the Faraday configuration. The device can selectively transmit one circular polarized light and reflect the opposite circular polarized light. (b) The negative refraction of TM-polarized light in a Weyl semimetal with the Voigt configuration. $\bm{S}_t$ is the transmitted Poynting vector. $\bm{k}_i$, $\bm{k}_r$, and $\bm{k}_t$ are the incident, reflected, and transmitted wavevectors, respectively. (c) A scheme of a scanning tunneling microscope (STM) with a probing tip built from three Weyl semimetal layers. These three layers act as an electronic Veselago lens. The electron flow from the top layer is focused on a small spot in the bottom layer. The tight focus may increase the spatial and temporal resolution of the STM. Figures are reproduced with permission from (a) Ref.~\cite{chtchelkatchev_chiral_2021}, Copyright 2021 Elsevier; (b) Ref.~\cite{ukhtary_negative_2017}, Copyright 2017 the Physical Society of Japan; (c) Ref.~\cite{hills_current-voltage_2017}, Copyright 2017 American Physical Society. }
    \label{fig:negative_refraction}
\end{figure}

Due to the giant gyrotropy of magnetic Weyl semimetals, the left and right circularly polarized light propagating along the chiral shift direction will acquire different phases and attenuation~\cite{kargarian2015}. Such effects can be used to construct polarization filters. 
In Ref.~\cite{chtchelkatchev_chiral_2021}, Chtchelkatchev et al.~demonstate that a single magnetic Weyl semimetal slab can selectively transmit/reflect circularly polarized light in the Faraday configuration, and linearly polarized light in the Voigt configuration (Fig.~\ref{fig:negative_refraction}a). In Ref.~\cite{yang_midinfrared_2022}, Yang et al.~design a circular polarizer using two layers of magnetic Weyl semimetals separated by an air gap. The chiral shifts of the two Weyl semimetals are parallel and perpendicular to the slab. The proposed device exhibits a high circular polarization efficiency and high average transmittance in the wavelength region from \SI{9}{\micro m} to \SI{15}{\micro m} at incidence angles up to \SI{50}{\degree}. In Ref.~\cite{mukherjee_absorption_2017}, Mukherjee et al.~study the effect of a tilt of the Weyl cones on the
absorption of left and right circular polarized light. They show that the difference in absorption depends strongly on the degree of tilt.

\subsubsection{Negative refraction}

Negative refraction refers to the counter-intuitive phenomenon where light is refracted at a negative angle with respect to the surface normal~\cite{veselago_electrodynamics_1968}. Negative refraction is related to exotic phenomena such as perfect lensing~\cite{pendry_negative_2000}.  
There are several known routes to achieving negative refraction. The original approach proposed by Veselago~\cite{veselago_electrodynamics_1968} requires the use of a negative index material where both the permittivity and the permeability are negative. Such ``double-negative" materials have been experimentally demonstrated by metamaterials~\cite{shelby_experimental_2001}. Negative refraction has also been achieved in photonic crystals~\cite{notomi2000,luo2002,cubukcu2003a}. Another approach utilizes the magnetoelectric effect in chiral media~\cite{tretyakov_waves_2003,pendry_chiral_2004,monzon2005}, which has also been experimentally confirmed using chiral metamaterials~\cite{zhang2009a}. 

Weyl semimetals can provide a new route to negative refraction using natural materials. As we discussed in Sec.~\ref{subsec:EM_wave_in_bulk}, TM-polarized light can propagate in a magnetic Weyl semimetal with the Voigt configuration when $\omega_{p-}<\omega<\omega_{p}$ or $\omega>\omega_{p+}$. In the lower frequency range, the light can propagate in the bulk below the plasma frequency $\omega_p$. This phenomenon is typical for any gyrotropic system in the Voigt configuration~\cite{huang2008,kotov_giant_2018} and is related to the modification of $\varepsilon_V$ by the gyrotropy as shown in Fig.~\ref{fig:scheme}c. Interestingly, this phenomenon is also accompanied by negative refraction. In Refs.~\cite{ukhtary_negative_2017,hayata_new_2018}, it is shown that TM-polarized light in the lower frequency range propagates with opposite signs of phase and group velocities. Hence the light exhibits a negative refractive index (Fig.~\ref{fig:negative_refraction}b). Although such negative refraction can be observed in any gyrotropic media, magnetic Weyl semimetals can exhibit the phenomena over a much broader bandwidth without any external magnetic field. These predictions still await experimental confirmation. 

Although not the focus of this paper, we point out that similar negative refraction for \emph{electrons} can also occur in a Weyl semimetal~\cite{hills_current-voltage_2017, tchoumakov_threedimensional_2022}. The phenomenon of negative refraction for electrons is similar to that for photons, although the physical mechanisms are different. Such an effect can be used to construct a three-dimensional electronic Veselago lens. In Ref.~\cite{hills_current-voltage_2017}, Hills et al.~design an electronic Veselago lens made of three layers of different Weyl semimetals. They propose to use such a lens as a probing tip for scanning tunneling microscope (STM) to significantly improve its spatial and temporal resolution (Fig.~\ref{fig:negative_refraction}c). In Ref.~\cite{tchoumakov_threedimensional_2022}, Tchoumakov et al. design a three-dimensional electronic Veselago lens made of a single Weyl semimetal. Such a Veselago lens is based on chiral anomaly and can selectively focus electrons of a given chirality.

\subsection{Nonlinear optical effects}\label{nonlinear}

So far, we have only discussed the linear optical effects of Weyl semimetals. These phenomena can be understood in the context of linear response theory~\cite{kubo1957,giuliani2005}, as we have demonstrated in Sec.~\ref{sec:optical_properties}. Weyl semimetals can also exhibit pronounced nonlinear optical effects~\cite{moore2019}. These phenomena must be understood in the context of nonlinear response theory~\cite{peterson1967}.  Nonlinear responses of topological materials have only been investigated recently~\cite{sodemann_quantum_2015,cortijo2016,morimoto2016b,morimoto_semiclassical_2016,ma_direct_2017,wu_giant_2017,de_juan_quantized_2017,chan_photocurrents_2017,konig_photogalvanic_2017,golub_photocurrent_2017}; a systematic discussion is beyond the scope of this paper. Instead, we survey some nonlinear optical effects and applications of Weyl semimetals.

\subsubsection{Photogalvanic effect}\label{photogalvanic}

One of the simplest nonlinear optical effects in a solid is the photogalvanic effect. It refers to the generation of the direct current (DC) in the crystal under exposure to light. In the context of nonlinear optics~\cite{boyd_nonlinear_2008}, the photogalvanic effect is a second-order nonlinear optical effect: A direct current can appear in a \emph{noncentrosymmetric} solid due to an oscillating electric field when one analyzes the response up to (at least) second order in the applied field~\cite{sturman1992}. At this order of perturbation theory, the DC photocurrents $\bm{J}$ are the sum of three contributions~\cite{sipe_second-order_2000,rostami_nonlinear_2018}:
\begin{equation}
    \bm{J} = \bm{J}_{\text{injection}} + \bm{J}_{\text{shift}} + \bm{J}_{\text{anomalous}},
\end{equation}
where the three terms are referred to as ``injection", ``shift", and ``anomalous" current contributions in the literature. These three contributions have different physical origins and symmetry requirements. Moreover, they exhibit different dependence on scattering time: $\bm{J}_{\text{injection}}$ is proportional to the scattering time $\tau$, while $\bm{J}_{\text{shift}}$ and $\bm{J}_{\text{anomalous}}$ have well-defined DC values even
in the absence of scattering~\cite{sipe_second-order_2000}. Roughly speaking,
\begin{equation}\label{eq:relation_scattering}
|\bm{J}_{\text{injection}}|/|\bm{J}_{\text{shift}}|\sim |\bm{J}_{\text{injection}}|/|\bm{J}_{\text{anomalous}}| \sim \omega \tau,
\end{equation}
where $\omega$ is the light frequency~\cite{rostami_nonlinear_2018}. Hence, the injection current dominates in the high-frequency or weak-scattering regime, while the shift or anomalous current dominates in the low-frequency or strong-scattering regime.

From a practical perspective, converting light to electricity is crucial for clean energy, imaging, communications, and chemical and biological sensing. For this purpose, the photogalvanic effect is of interest because it can potentially overcome the extrinsic limitations of conventional approaches. For example, traditional solar cells make use of the built-in electric fields in p-n junctions to generate photocurrent; however, their efficiency is bounded by the Shockley-Queisser limit due to the constraint of detailed balance~\cite{shockley1961}. Thermoelectric devices utilize the optically induced thermal gradients to produce currents via the Seebeck effect; however, they require a careful balance of the optical, electronic, and thermal material properties. The photogalvanic effect provides an important alternative: It has an ultrafast response with fewer limitations on efficiency or maximum open-circuit voltage~\cite{spanier2016}.

\begin{figure}[htbp]
    \centering
    \includegraphics[width=0.9\textwidth]{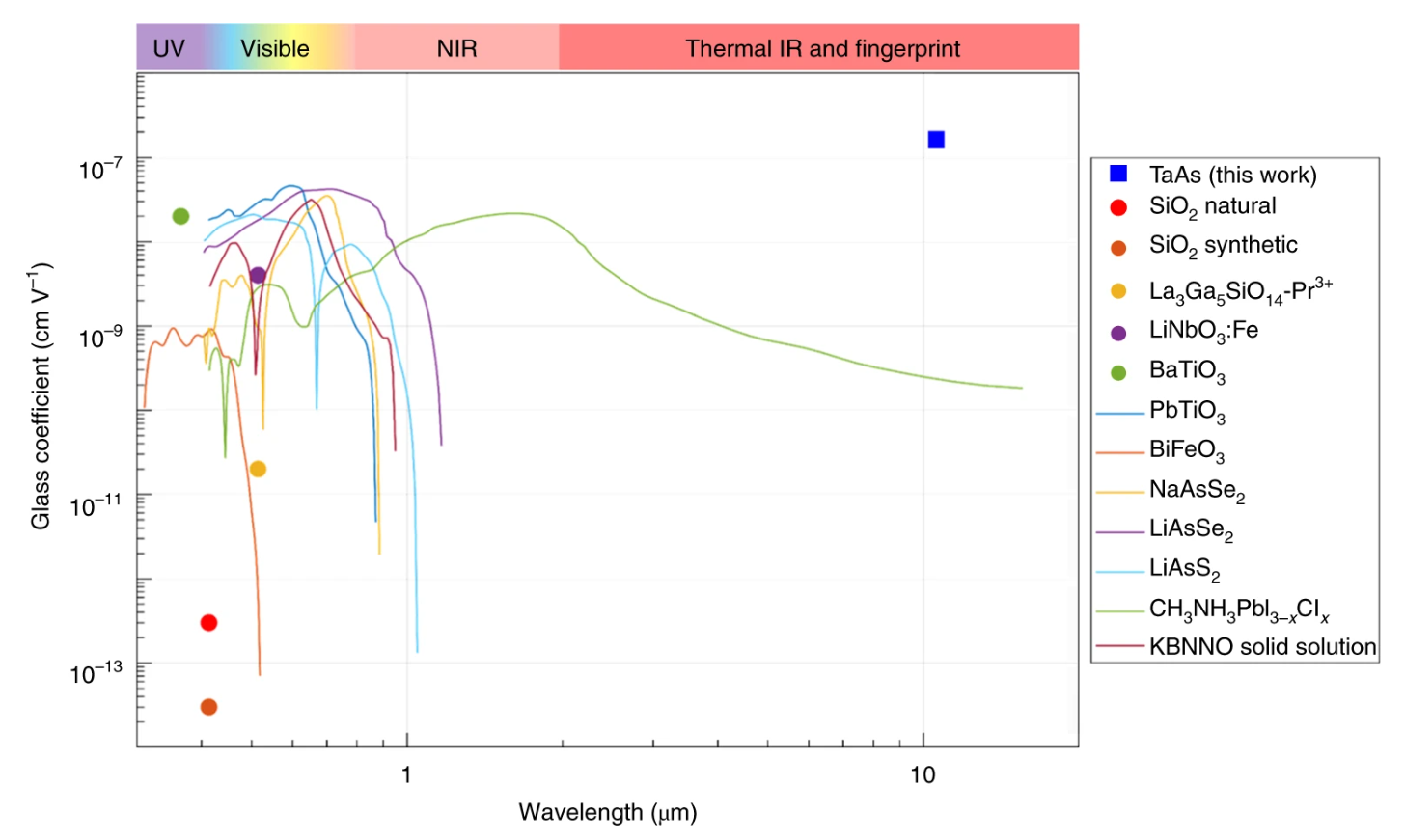}
    \caption{\csentence{Glass coefficients.} Measured (circles) and calculated (lines) Glass coefficients for various ferroelectric materials.  The Weyl semimetal TaAs (blue square) exhibits a nearly one order of magnitude larger response due to its anomalous Berry curvature. TaAs works in the technically important  mid-infrared regime due to its gapless nature. Figure is reproduced with permission from Ref.~\cite{osterhoudt_colossal_2019}. Copyright 2019 Springer Nature. }
    \label{fig:Glass}
\end{figure}

The photogalvanic effects were primarily observed in ferroelectric insulators and semiconductors~\cite{sturman1992}. However, in these materials, the photogalvanic effects are usually too weak to be technically relevant. Moreover, the photogalvanic effects are typically restricted to a narrow range of light wavelengths. For the practical usage of photogalvanic effects, it is crucial to search for materials that overcome these limitations.

Weyl semimetals can be ideal material candidates for this purpose. Recent studies show that noncentrosymmetric Weyl semimetals can exhibit a much stronger photogalvanic effect than conventional materials due to the large Berry curvature~\cite{chan_photocurrents_2017,osterhoudt_colossal_2019}. The operation is broadband since Weyl semimetals exhibit zero bandgap and linear dispersion~\cite{zhang_highfrequency_2021,wang_hybrid_2022}. Moreover, the carriers in Weyl semimetals can exhibit high mobility; the fast motion of carriers also contributes to the  ultrafast and giant photocurrent response~\cite{weng2019}. Interestingly, Weyl semimetals exhibit topological features in all three contributions of photocurrents~\cite{morimoto2016b}.

To compare the strength of photogalvanic effects in different materials, one calculates the so-called Glass coefficient $G$ defined by~\cite{glass1974,sturman1992}:
\begin{equation}\label{eq:Glass_coefficient}
    j = G \alpha I, 
\end{equation}
where $j$ is the photocurrent density
, $\alpha$ is the absorption coefficient
, and $I$ is the incident intensity
. Thus $G$ measures the generated photocurrent per absorbed power with a unit $[\si{cm/V}]$. Fig.~\ref{fig:Glass} shows the measured and calculated Glass coefficients for various ferroelectric materials, together with the measured data for the Weyl semimetal TaAs. Notably, TaAs exhibits a nearly one order of magnitude larger response than any other materials. Moreover, TaAs enables a broadband response even in the mid-IR wavelengths, a regime that is important for thermal, chemical, and biological sensing.

Below we discuss the three contributions of photocurrents in Weyl semimetals. For each contribution, we first provide a qualitative picture of the process in general solids to explain its physical origin and symmetry requirement. Then we discuss its manifestation in Weyl semimetals. We point to relevant references for detailed quantitative analysis. 

\begin{figure}[htbp]
    \centering
    \includegraphics[width=0.80\textwidth]{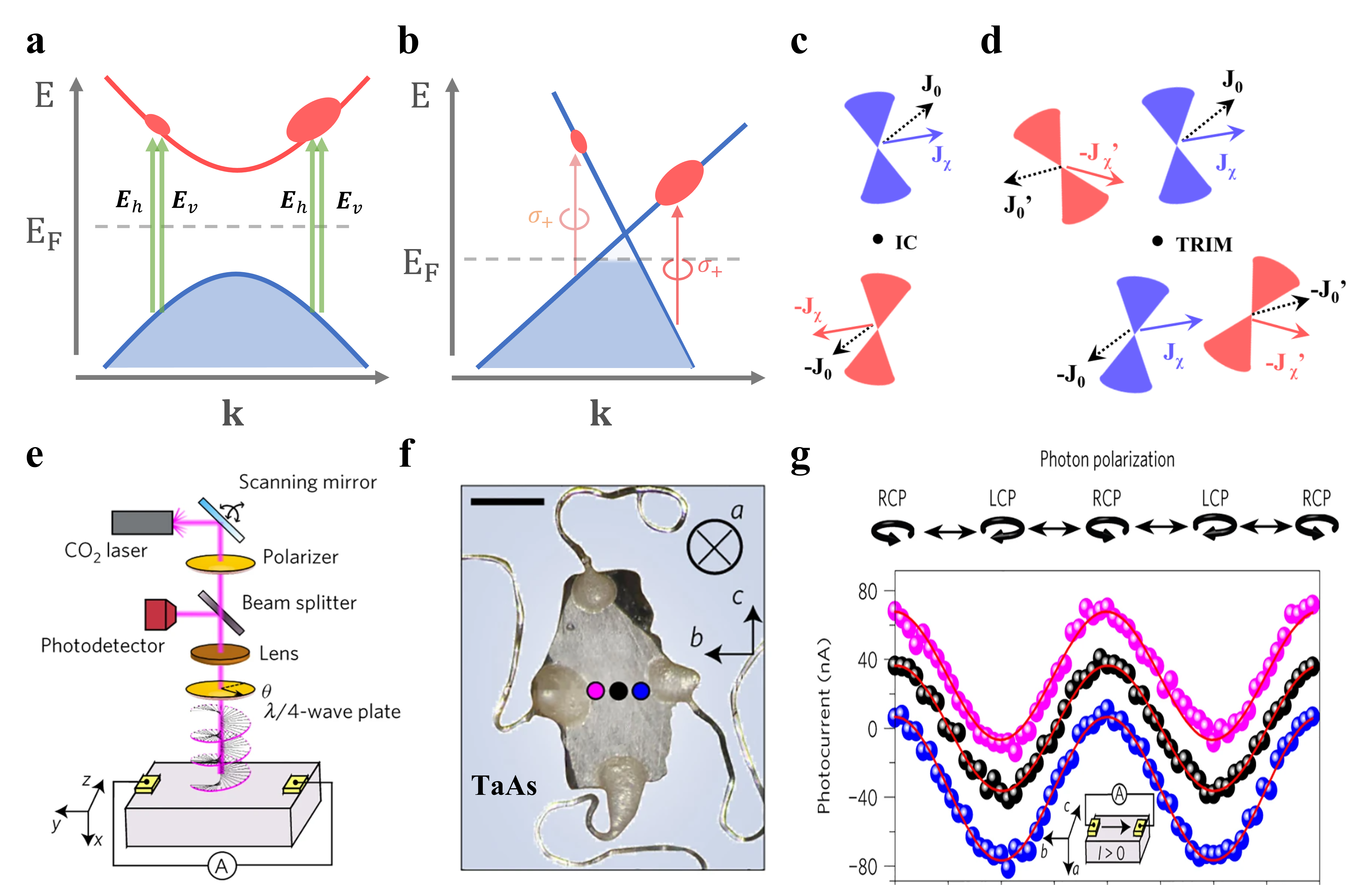}
    \caption{\csentence{Injection current.} (a) Scheme of injection current via interband absorption involving different polarization component $\bm{E}_h$ (horizontal) and $\bm{E}_v$ (vertical). (b) The injection current from a single Weyl cone depends on its chirality and tilt. (c, d) Injection currents in Weyl semimetals. Each Weyl node produces an injection current with chirality-independent ($\bm{J}_0$) and chirality-dependent ($\bm{J}_\chi$) components. (c) Inversion symmetry relates two Weyl nodes with opposite tilt and opposite chirality, leading to the cancellation of photocurrents. IC: the inversion center. (d) Time-reversal symmetry relates two Weyl nodes with opposite tilt but the same chirality.     Two pairs of Weyl nodes give rise to an overall injection current $2(\bm{J}_\chi - \bm{J}'_\chi)$. TRIM: the time-reversal invariant momentum. (e-g) Experimental observation of the injection current in the Weyl semimetal TaAs. (e) Setup scheme. (f) A photograph of the sample. $\bm{a}, \bm{b}, \bm{c}$ denote the crystal axes. Scale bar: \SI{300}{\micro m}. (g) Polarization-dependent photocurrents at $T=\SI{10}{K}$ measured along the $\bm{b}$ direction with the laser focused on the pink, black and blue dots in (f). Figures are reproduced with permission from (c, d) Ref.~\cite{chan_photocurrents_2017}, Copyright 2017 American Physical Society; (e-g) Ref.~\cite{ma_direct_2017}, Copyright 2017 Springer Nature. }
    \label{fig:injection_current}
\end{figure}

First, we discuss the injection current. Injection currents arise in a crystal that lacks certain symmetries, which give rise to phase differences between transition amplitudes associated with different polarizations of light~\cite{nastos2006}. When the crystal is excited with elliptically polarized light, the different excitation pathways for horizontal and vertical linear polarizations lead to an interference effect. This results in an asymmetric population in the wavevector space and thus generates a current (Fig.~\ref{fig:injection_current}a). The current flips sign when the helicity of light changes sign. Because it is generated only with elliptically polarized light, the injection current is also called ``circular photocurrent"~\cite{laman1999}.

The injection current requires the breaking of inversion symmetry. But this is not sufficient. Detailed symmetry analysis~\cite{sipe_second-order_2000} shows that among the $21$ crystal classes that lack inversion symmetry, $18$ classes can support injection current; the exceptions are $\bar{6}m2$, $\bar{6}$, and $\bar{4}\bar{3}m$~\cite{sipe_second-order_2000}. For example, GaAs belongs to $\bar{4}\bar{3}m$, hence cannot support injection current.

In a Weyl semimetal, the injection current can occur via asymmetric interband excitations near each Weyl cone (Fig.~\ref{fig:injection_current}b). Each Weyl node produces an injection current that depends on its chirality and tilt, which can be decomposed as~\cite{chan_photocurrents_2017}:
\begin{equation}\label{eq:injection_Weyl_components}
    \bm{J}^{(n)} = \bm{J}^{(n)}_0 + \bm{J}^{(n)}_\chi, 
\end{equation}
where $(n)$ denotes the $n$-th Weyl cone, $\bm{J}^{(n)}_0$ changes sign only for opposite tilts, and $\bm{J}^{(n)}_\chi$ changes sign only for opposite chirality. The total injection current is then the sum of contributions from all the Weyl cones. 

Next, we discuss the symmetry constraints. When there is inversion symmetry, a Weyl node at $\bm{k}$ is related to another one at $-\bm{k}$ with opposite tilt and opposite chirality. When there is time-reversal symmetry, a Weyl node at $\bm{k}$ is related to another one at $-\bm{k}$ with opposite tilt but the same chirality. Combining these symmetry considerations with Eq.~(\ref{eq:injection_Weyl_components}), we conclude that in centrosymmetric Weyl semimetals, both $\bm{J}_0$ and $\bm{J}_\chi$ are canceled; the total injection current is zero as expected (Fig.~\ref{fig:injection_current}c). In noncentrosymmetric Weyl semimetals with time-reversal symmetry, $\bm{J}_0$ are canceled while $\bm{J}_\chi$ can survive; a minimal number of two pairs of Weyl nodes can produce a nonzero response $2(\bm{J}_\chi - \bm{J}'_\chi)$  (Fig.~\ref{fig:injection_current}d). In Weyl semimetals without any symmetry, both $\bm{J}_0$ and $\bm{J}_\chi$ can survive~\cite{chan_photocurrents_2017}. These discussions confirm the general symmetry analysis above from a different microscopic perspective.

The injection current in Weyl semimetals has been investigated both theoretically~\cite{morimoto_semiclassical_2016,chan_photocurrents_2017,de_juan_quantized_2017,golub_photocurrent_2017,zhang_photogalvanic_2018} and experimentally~\cite{ma_direct_2017,sun_circular_2017, ji_spatially_2019}. In Ref.~\cite{ma_direct_2017}, Ma et al.~report the experimental observation of the injection current in TaAs. They use a mid-infrared scanning photocurrent microscope (Fig.~\ref{fig:injection_current}e) equipped with a CO$_2$ laser with wavelength $\lambda_{\text{CO}_2} = \SI{10.6}{\micro m}$ and energy $\hbar \omega = \SI{117}{meV}$. This photon energy is chosen to induce the desired interband transitions near the Weyl nodes. The TaAs sample is purposedly filed down to show a clean surface with its normal direction along the $\bm{a}$ ($[100]$) axis of the crystal (Fig.~\ref{fig:injection_current}f). The sample is kept at a low temperature $T\approx \SI{10}{K}$ to increase the scattering time $\tau$, which is crucial to observe the shift current [see  Eq.~(\ref{eq:relation_scattering})]. Light is focused and normally incident on the sample, i.e.,~along the $\bm{a}$ axis. Its polarization is controlled by the rotation angle $\theta$ of the quarter-wave plate. In  Fig.~\ref{fig:injection_current}g, the pink, black, and blue data points show the photocurrent along the $\bm{b}$ axis when the laser is focused on the pink, black, and blue dots in Fig.~\ref{fig:injection_current}f, respectively. When the laser is focused near the sample's center (the black dot), the photocurrent fits well to a cosine function of $\theta$: It reaches the maximum for right circularly polarized light, the minimum for left circularly polarized light, and zero for linearly polarized light. When the laser spot is moved horizontally to the blue and pink dots, the corresponding photocurrents exhibit the same polarization dependence but with an additional, polarization-independent shift. These data reveal two distinct mechanisms for photocurrent generation. The polarization-dependent component corresponds to the injection current. The polarization-independent component arises from the photo-thermal effect~\cite{gabor2011,mciver2012,yuan2014}: Because the sample and contact have different thermopower, a current is generated by the laser-induced temperature gradient. Such an interpretation is further supported by additional experiments~\cite{ma_direct_2017}.

Second, we discuss the shift current. Shift currents arise from a coordinate shift accompanying the interband photoexcitation of electrons~\cite{tan2016}. It occurs because the real-space center of charge for the valence bands differs from that for the conduction bands. As light is absorbed and electrons transit from the valence to conduction bands, there will be a motion of charge. If the crystal has low enough symmetry and the light polarization is appropriate, there will be a net current due to the ``shift" of the center of charge~\cite{nastos2006}. The shift is on the order of a bondlength, and occurs on femtosecond time scales~\cite{sipe_second-order_2000}. A more rigorous treatment of the shift current requires the modern theory of electric polarization~\cite{vanderbilt2018a} and the formalism of localized Wannier functions~\cite{ibanez-azpiroz2018,lihm2021}. 

\begin{figure}[htbp]
    \centering
    \includegraphics[width=0.9\textwidth]{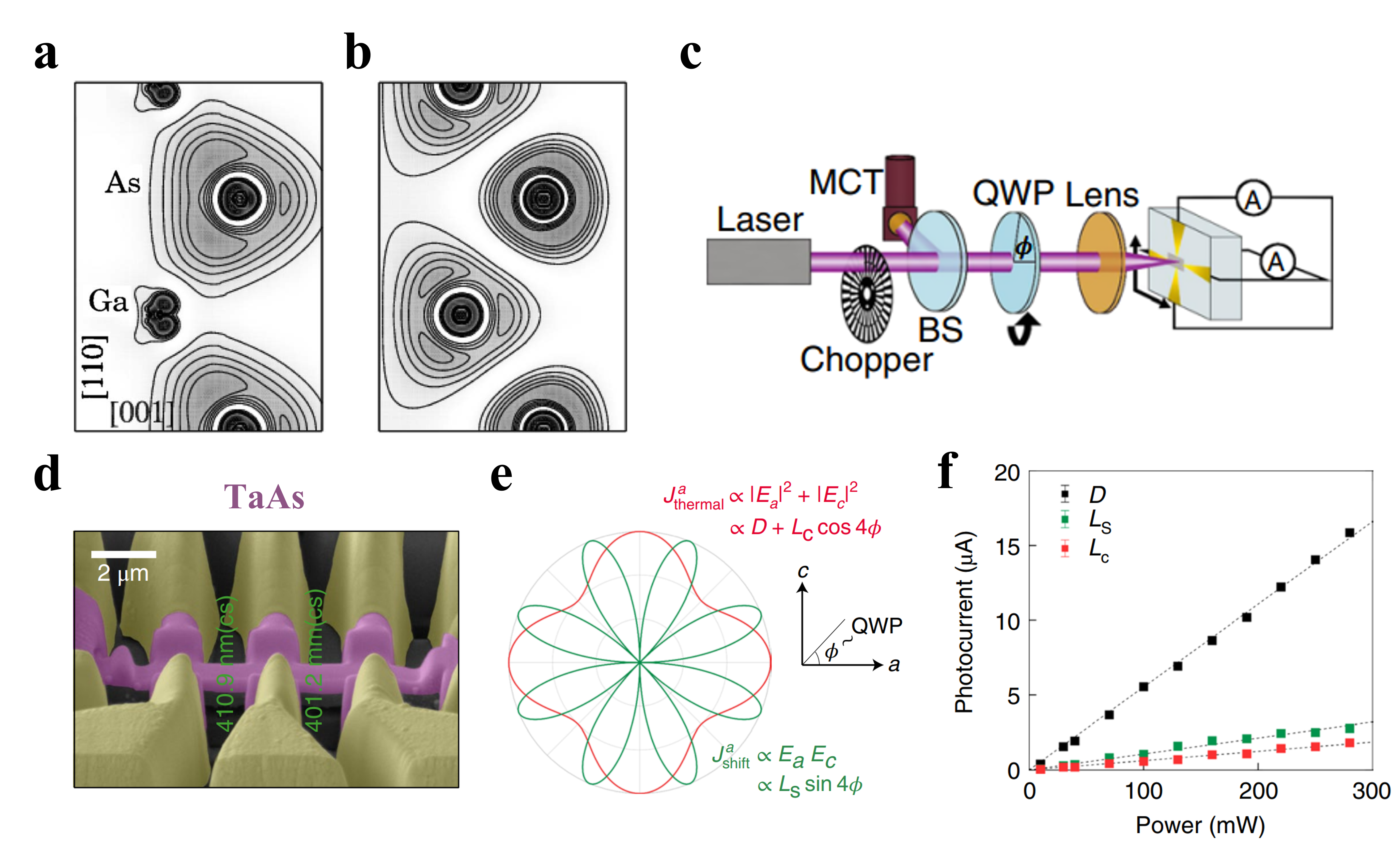}
    \caption{\csentence{Shift current.} (a, b) Electron density in the $[110]$ plane of GaAs at $\Gamma$ point for (a) the highest valence band, and (b) the lowest conduction band. (c-f) Experimental observation of the shift current in the Weyl semimetal TaAs. (c) Setup scheme. The polarization is controlled via rotation of the quarter-wave plate (QWP). (d) False colored scanning electron microscopy (SEM) image of a microscopic TaAs (purple) device with gold (yellow) contacts. (e) Polarization dependence of the thermal ($J^{a}_{\text{thermal}}$) and shift ($J^{a}_{\text{shift}}$) photocurrent contributions. The radius and angle of the polar plot correspond to the magnitude of the photocurrents and the angle ($\phi$) the fast axis of QWP makes with the crystal $a$ axis, respectively. These responses exhibit out-of-phase minima and maxima due to the different electric field combinations. (f) Measured power dependence of QWP angle-independent ($D$), fourfold shift ($L_s$), and photothermal ($L_c$) terms. The dependence is linear as expected from the generation mechanisms of the shift and photothermal current. Figures are reproduced with permission from (a, b)  Ref.~\cite{nastos2006}, Copyright 2006 American Physical Society; (c-f) Ref.~\cite{osterhoudt_colossal_2019}, Copyright 2019 Springer Nature. }
    \label{fig:shift_current}
\end{figure}

As an illustration, we discuss the shift current in the intrinsic semiconductor GaAs~\cite{nastos2006}. Fig.~\ref{fig:shift_current}a and~\ref{fig:shift_current}b show the electron density of GaAs for the $\Gamma$ electrons in the valence and conduction bands, respectively. They can be viewed as snapshots before and after the photoexcitation. Electrons at the top of the valence band ($\Gamma$ point) are localized around the As atoms (Fig.~\ref{fig:shift_current}a). During the photoexcitation, the crystal absorbs photons of energy larger than the bandgap to populate the states near the bottom of the conduction band (also $\Gamma$ point). Now electrons have relocated closer to the Ga atoms (Fig.~\ref{fig:shift_current}b). Depending on the light polarization, the electron density evolves differently to reach the excited state. For example, if the electric field is polarized along the $[100]$ direction, an electron from the As atom can move towards any one of its four nearest neighboring Ga atoms with equal probability, generating no net current. If the polarization is along the $[111]$ direction, an electron from the As atom will move primarily towards its closest neighboring Ga atom in the $[111]$ direction, generating a net current~\cite{nastos2006}.

The shift current requires the breaking of inversion symmetry. But, again, this is not sufficient. Detailed symmetry analysis~\cite{sipe_second-order_2000} shows that among the $21$ crystal classes that lack inversion symmetry, $20$ classes can support the shift current; the only exception is $432$. Thus the shift current exists in GaAs and other zinc-blende ($\bar{4}\bar{3}m$) semiconductors~\cite{cote2002}. Different from the injection current, the shift current can be generated with linearly polarized light.

The shift current in Weyl semimetals has been investigated both theoretically~\cite{morimoto2016b,zhang_photogalvanic_2018} and experimentally~\cite{osterhoudt_colossal_2019,ma_nonlinear_2019}. In Ref.~\cite{osterhoudt_colossal_2019}, Osterhoudt et al.~report the experimental observation of the shift current in the Weyl semimetal TaAs. Their setup is similar to that in Fig.~\ref{fig:injection_current}e: a photocurrent microscope equipped with a CO$_2$ laser, where the light polarization is controlled by the rotation angle $\phi$ of the quarter-wave plate (Fig.~\ref{fig:shift_current}c). Their sample is different from the bulk crystal in Fig.~\ref{fig:injection_current}f: The TaAs sample is fabricated into a microscopic device (Fig.~\ref{fig:shift_current}d) with a small area ($20$ times smaller than the laser spot) and a small thickness (only three times the penetration depth $\sim\SI{250}{nm}$ of the light). This design mitigates the resistive losses and thermal effects. The sample is kept at room temperature, hence the scattering time $\tau$ is short, and the shift current dominates over the injection current [see Eq.~(\ref{eq:relation_scattering})]. Despite the careful design of the device, the photo-thermal effect cannot be eliminated. Fortunately, one can separate the photo-thermal effect from the shift current by their different polarization dependence. Detailed symmetry analysis shows that for the photocurrent measured along the $a$ axis, the thermal contribution $J^{a}_{\text{thermal}} = D + L_c \cos 4\phi$  and the shift contribution $J^{a}_{\text{shift}} = L_s \sin 4\phi$ (Fig.~\ref{fig:shift_current}e). By fitting the $\phi$-dependent photocurrent, one can obtain the parameters $D$, $L_c$, and $L_s$. Fig.~\ref{fig:shift_current}f shows their measured values as functions of input power, which all exhibit linear dependence. This is as expected since both mechanisms of the shift current and the photo-thermal effect rely on the square of the electric field.

Third, we briefly discuss the anomalous current. The anomalous current arises from the anomalous velocity~\cite{karplus1954,sundaram1999} caused by the Berry curvature~\cite{vanderbilt2018a,sodemann_quantum_2015, zhang_berry_2018,gao_second-order_2020}. Similar to the injection current, the anomalous current is generated only with elliptically polarized light, and changes sign when the helicity of light flips~\cite{rostami_nonlinear_2018}. The anomalous current in Weyl semimetals was theoretically predicted but not experimentally observed yet. We refer readers to Ref.~\cite{rostami_nonlinear_2018} for detailed theoretical analysis.

Finally, we point out that the Fermi arc states may induce additional contribution to the photocurrent generation. See Refs.~\cite{chang_unconventional_2020,steiner_surface_2022} for more details.

\subsubsection{Orbital angular momentum detection}\label{OAMdet}

\begin{figure}[htbp]
    \centering
    \includegraphics[width=0.95\textwidth]{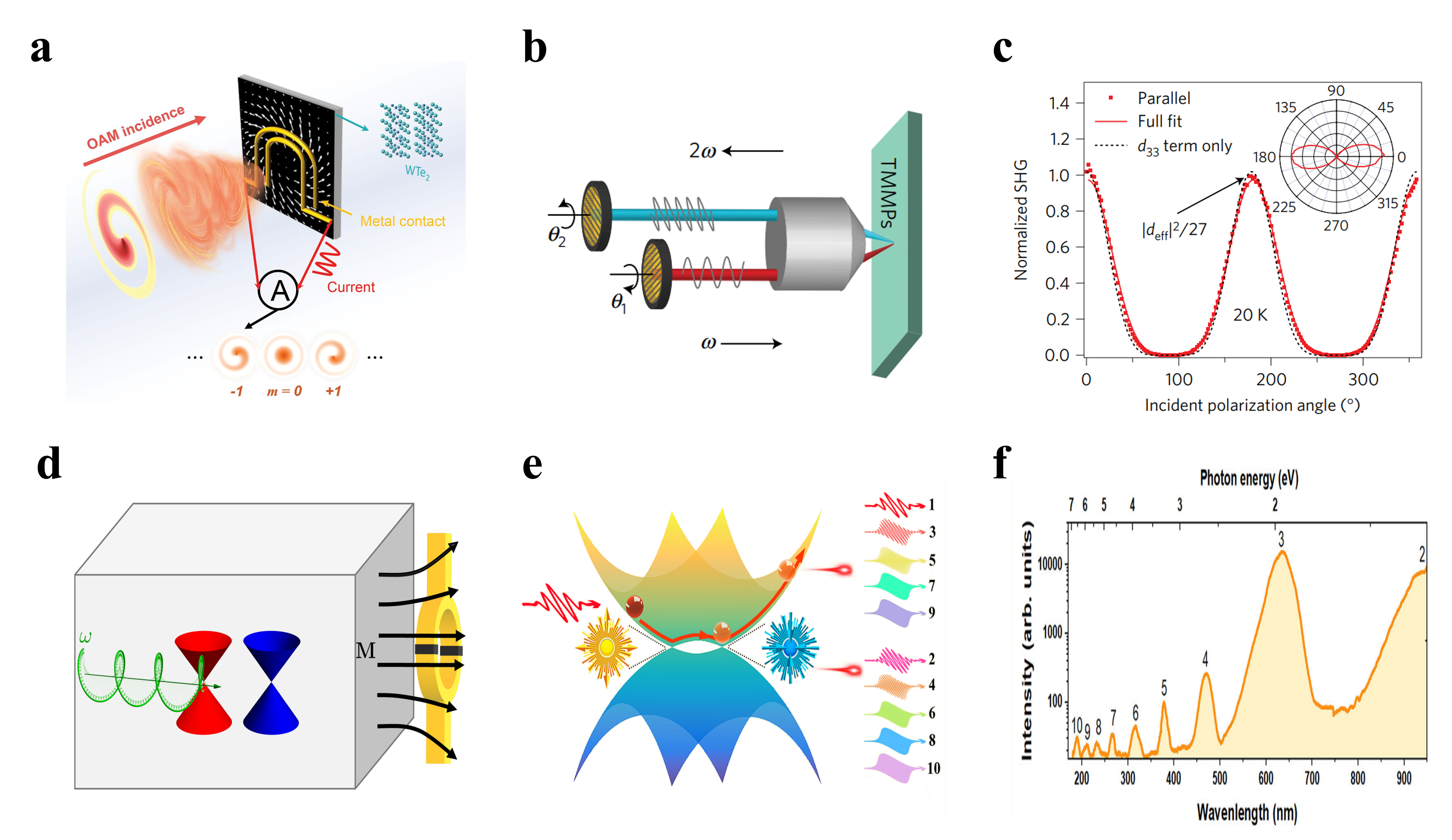}
    \caption{\csentence{Some nonlinear optical effects in Weyl semimetals.}  (a) Direct detection of light orbital angular momentum. (b, c) Second harmonic generation. (b) Setup scheme. (c) Second harmonic signal as a function of incident polarization angle. (d) Inverse Faraday effect. (e, f) High harmonic generation. (e) Scheme. (f) High harmonic spectrum. Figures are reproduced with permission from (a) Ref.~\cite{ji_photocurrent_2020}, Copyright 2020 American Association for the Advancement of Science; (b, c) Ref.~\cite{wu_giant_2017}, Copyright 2017 Springer Nature; (d) Ref.~\cite{liang2021b}; (e, f) Ref.~\cite{lv_highharmonic_2021}.}
    \label{fig:other_nonlinear_effects}
\end{figure}

Light can carry orbital angular momentum (OAM)~\cite{allen_orbital_1992}. The OAM manifests as a helical wavefront with an azimuthal phase distribution $e^{im\phi}$ where $m$ is the mode number and $\phi$ is the azimuthal angle. The OAM modes of light can encode information. However, direct detection of OAM by the photocurrent mesurement is challenging. This is because most types of photocurrents are sensitive only to optical intensity, not to optical phase. 

In Ref.~\cite{quinteiro2009}, Quienteiro et al.~propose a new mechanism for the generation of photocurrent, called the orbital photogalvanic effect. In this case, the incident light can transfer its OAM and energy simultaneously to the electrons. This process is similar to the photon drag effect~\cite{lebedew1901}, where the linear momentum of absorbed photons is transferred to electrons. Because the optical phase varies in the azimuthal direction, it induces a spatial imbalance of excited carriers, producing a net current flowing either along or perpendicular to the helical phase gradient. The generated photocurrent is proportional to the OAM; when the OAM reverses signs, the photocurrent also flips direction. 

In Ref.~\cite{ji_photocurrent_2020}, Ji et al.~utilize the orbital photogalvanic effect to enable direct on-chip electric readout of OAM. They fabricate electrodes of various shapes on WTe$_2$ for use as photocurrent detectors (Fig.~\ref{fig:other_nonlinear_effects}a). WTe$_2$ is a room-temperature Weyl semimetal with broken inversion symmetry. It is an ideal material for observing the orbital photocurrent because it has large nonlinear optical susceptibilities and certain symmetries that forbid the photocurrents of other types. In the experiment, Ji et al.~observe that the photocurrent displays steplike changes with different OAM. Such a detector can be exploited in future photonic circuits for optical communications.  In Ref.~\cite{lai_direct_2022}, Lai et al.~demonstrate the generation of orbital photocurrents in the mid-infrared wavelengths with another Weyl semimetal TaIrTe$_4$.

\subsubsection{Second harmonic generation}\label{subsec:SHG}

Second harmonic generation is another important second-order nonlinear optical effect. It refers to the generation of photons with twice the frequency of the incident light. Recent works~\cite{wu_giant_2017,patankar_resonance-enhanced_2018} show that noncentrosymmetric Weyl semimetals can exhibit strong second harmonic generation. 

In Ref.~\cite{wu_giant_2017}, Wu et al.~reveal giant second harmonic generation in the Weyl semimetals TaAs, TaP, and NbAs. These materials belong to the point group $4mm$, which has a unique polar ($z$) axis along the $[001]$ direction. Symmetry analysis shows that second harmonic generation is allowed when the incident electric field has a nonzero $z$-component. Fig.~\ref{fig:other_nonlinear_effects}b depicts the schematic setup. The incident pulses of $\SI{800}{nm}$ wavelength are focused at near-normal incidence on the $(112)$ surface of the sample. The generated second-harmonic light is detected in the reflection. The polarization of the incident light and the second-harmonic light are controlled by two sets of polarizers and waveplates, referred to as the generator and the analyzer, respectively. Angles $\theta_1$ and $\theta_2$ denote the angles of the linear polarization plane after the generator and the analyzer, respectively, with respect to the $[1,1,-1]$ crystal axis. Fig.~\ref{fig:other_nonlinear_effects}c shows the normalized second harmonic intensity as a function of $\theta_1$ when $\theta_1 = \theta_2$ at $\SI{20}{K}$, while the inset shows the polar plot. The data is consistent with the symmetry analysis and can be fitted well with the theoretical prediction.  The second harmonic signal is highly anisotropic and sensitive to the incident polarization: It reaches the maximum at $\theta_1 = \SI{0}{\degree}$ or $\SI{180}{\degree}$ when the electric field is along the $[1,1,-1]$ direction and has the largest $z$-component; it reaches the minimum at $\theta_1 = \SI{90}{\degree}$ or $\SI{270}{\degree}$ when the electric field is along the $[1,-1,0]$ direction and has zero $z$-component. The second harmonic signal is giant: The second-order optical susceptibility $\chi^{(2)}$ has the largest component of $\SI{7200}{pm V^{-1}}$, which is larger by almost one order of magnitude than the value in the archetypal electro-optic materials GaAs ($\SI{700}{pm V^{-1}}$) and ZnTe ($\SI{900}{pm V^{-1}}$), even when measured at wavelengths where their response is the largest. In fact, such a value is larger than then reported in any crystal. In Ref.~\cite{patankar_resonance-enhanced_2018}, Patankar et al.~further investigate the physical origins of such a giant second harmonic generation in TaAs.

From a practical perspective, Weyl semimetals are not optimal for frequency-doubling applications in the visible regime because of their strong absorption. However, they are promising materials for terahertz generation~\cite{gao2020f} and optoelectronic devices such as far-infrared detectors because the loss is lower at these longer wavelengths.

\subsubsection{Inverse Faraday effect}\label{subsubsec:inverse_faraday}

The inverse Faraday effect is the reverse of the Faraday effect. Recall that the Faraday effect refers to the polarization rotation of light induced by a static magnetization along the light propagation direction~\cite{zvezdin_modern_1997}. Correspondingly, the inverse Faraday effect refers to the generation of static magnetization induced by circularly polarized light propagating inside the material (Fig.~\ref{fig:other_nonlinear_effects}d)~\cite{van_der_ziel_optically-induced_1965,hertel_theory_2006}. In this case, the angular momentum from an oscillating electric field $\bm{E}(t)$ is transferred into the magnetic moment of electrons, leading to a static magnetization $\bm{M} \propto \bm{E}(\omega)\times\bm{E}^*(\omega)$. The inverse Faraday effect is a nonlinear phenomenon~\cite{pershan_nonlinear_1963}. It does not require absorption but is rather based on a Raman-like coherent optical scattering process. Consequently, this effect is non-thermal and takes place on a femtosecond timescale. The laser-induced magnetization can be used for ultrafast control of magnetic devices~\cite{kimel_ultrafast_2005,kimel_femtosecond_2007}. It can also cause the Faraday rotation for a subsequent \emph{probe} light, which induces a matter-mediated light-light interaction.

Recent works~\cite{zyuzin_nonlinear_2018,kawaguchi_giant_2020,tokman_inverse_2020,gao_topological_2020,cao_lowfrequency_2022} show that Dirac and Weyl semimetals can exhibit a strong inverse Faraday effect due to their unique carrier transport properties and strong spin-orbit coupling. For example, in Ref.~\cite{tokman_inverse_2020}, Tokman et al.~estimate that in a Weyl semimetal with $n$ Weyl nodes and a Fermi energy of $\SI{100}{meV}$, an incident pump light with an electric field  $\SI{10}{kV/cm}$ at a frequency $\SI{1}{THz}$ can induce a magnetization that causes the Faraday rotation parameter $\alpha \approx 6.6  n^{3/2}$~\si{rad/cm} for a probe light. Such a value is two orders of magnitude larger than that obtained in typical ferrites~\cite{kimel_femtosecond_2007}.

In Ref.~\cite{liang2021b}, Liang et al. propose an unconventional mechanism of inverse Faraday effect in Dirac and Weyl semimetals, referred to as the ``axial magnetoelectric effect". As we discussed above, in the conventional inverse Faraday effect, 
static magnetization is generated through dynamical electromagnetic fields. In contrast, in the axial magnetoelectric effect, static magnetization is generated through dynamical axial gauge fields. Such effective gauge fields can be generated in a Dirac or Weyl semimetal via dynamical deformations (sound). They interact with electrons similarly to the usual electromagnetic fields but with different signs for different chiralities (see Ref.~\cite{gorbar2021} for an introduction to axial gauge fields). In the axial magnetoelectric effect, 
the angular momentum is transferred from the axial or pseudoelectric field, conventionally denoted as  $\bm{E}_5(t)$, into the magnetic moment of electrons, leading to a static magnetization $\bm{M}\propto \bm{E}_5(\omega)\times \bm{E}_5^*(\omega)$. The synthetic nature of the axial gauge fields  means that one can induce magnetization in Dirac and Weyl semimetals using phonons without any electromagnetic fields.

\subsubsection{Higher-order nonlinear effects}\label{subsubsec:high-order-nonlinear}

Weyl semimetals can also exhibit higher-order nonlinear optical effects, such as four-wave mixing, optical Kerr effect, and high-harmonic generation. These nonlinear effects are expected to be strong, especially at long wavelengths, due to the linear dispersion and high mobility of the Weyl fermions. Moreover, these effects may exhibit nontrivial features due to the anomalous Berry curvature associated with the Weyl nodes.

In Ref.~\cite{almutairi_four-wave_2020}, Almutairi et al.~derive the third-order nonlinear optical conductivity of Weyl semimetals in the long-wavelength limit and calculate the intensity of the nonlinear four-wave mixing signal. The calculated nonlinear generation efficiency is surprisingly high for a lossy material, of the order of several $\si{mW}$ per $\si{W^3}$ of the incident pump power. This value is many orders of magnitude higher than in conventional nonlinear materials~\cite{boyd_nonlinear_2008}. Optimal conditions for the four-wave mixing are realized in the vicinity of bulk plasma resonance. This work indicates that ultrathin Weyl semimetal films of the order of skin depth in thickness can find applications in   compact nonlinear optoelectronic devices. 

In Ref.~\cite{cheng_third-order_2020}, Cheng et al.~derive the analytic expressions for linear and third-order optical conductivities of Dirac and Weyl semimetals, and compare the results with those of two-dimensional Dirac materials such as graphene. The details of the third-order conductivity are discussed for third-harmonic generation, the Kerr effect and two-photon carrier injection, parametric frequency conversion, and two-color coherent current injection. In contrast with two-dimensional materials, the three-dimensional Dirac and Weyl semimetals allow for adjusting the nonlinear signals by tuning the sample thickness. Thus, one can envision broad applications of such materials in nonlinear photonic devices.

In Ref.~\cite{lv_highharmonic_2021}, Lv et al.~report the experimental observation of high-harmonic generation in Weyl semimetal $\beta$-WP$_2$. High-harmonic generation in solids has only been discovered recently~\cite{ghimire2011,liu2017g,liu2018c,goulielmakis2022}. Recent works~\cite{bai2021a,heide2022} report efficient high-harmonic generation from topological materials. Moreover, Ref.~\cite{luu2018} shows that the polarimetry of high-harmonic emission from solids can be used to directly retrieve the Berry curvature. In Lv's experiment, both odd and even orders of high-harmonic emissions are observed (Fig.~\ref{fig:other_nonlinear_effects}f). The high-harmonic spectrum extends into the vacuum ultraviolet region (\SI{190}{nm}, 10th order) even under fairly low femtosecond laser intensity ($\sim\SI{0.29}{TW/cm^2}$). It is interpreted that odd-order harmonics come from the Bloch oscillation, while even-order harmonics arise from the effective Lorentz force due to the Berry curvature (Fig.~\ref{fig:other_nonlinear_effects}e). By analyzing the crystallographic orientation-dependent high-harmonic spectra, they further quantitatively retrieve the electronic band structure and Berry curvature of $\beta$-WP$_2$.

\subsection{Other effects}\label{subsec:other_effects}

Of course, this tutorial review cannot cover all the reported optical phenomena and applications of Weyl semimetals. Here we list some examples of other phenomena and applications not covered in this review: gyrotropic magnetic effect~\cite{zhong_gyrotropic_2016},
photoinduced anomalous Hall effect~\cite{chan_when_2016},
nonlinear magneto-optical effects~\cite{morimoto_semiclassical_2016}, 
topological frequency conversion~\cite{nathan_topological_2022},
emergent electromagnetic induction~\cite{ishizuka_emergent_2016,ishizuka_momentum-space_2017}, tunable perfect absorption~\cite{halterman_epsilon-near-zero_2018}, unidirectional bound states in the continuum~\cite{zhao_unidirectional_2022}, transverse Kerker effect~\cite{liu_transverse_2021}, and topological lasers~\cite{oktay_lasing_2020}.

\section{Thermal photonic  applications and devices}\label{sec:thermal_dev}

In this section, we survey various thermal photonic applications of Weyl semimetals. We discuss the nonreciprocal thermal emitters in Sec.~\ref{subsec:thermal_emitter}, the heat flux control in Sec.~\ref{subsec:heat_flux_control}, and the control of Casimir force in Sec.~\ref{subsec:Casimir_force_control}. 

\subsection{Nonreciprocal thermal emitters}\label{subsec:thermal_emitter}

Conventional thermal emitters obey Kirchhoff’s law of thermal radiation~\cite{kirchhoff1860,chen2005,zhang2007,bergman2011fundamentals,modest2021radiative,howell2020thermal}, which states that for a given direction, polarization, and frequency, the emissivity and the absorptivity are equal. However, Kirchhoff’s law is not a requirement of thermodynamics but a consequence of Lorentz reciprocity~\cite{snyder1998thermodynamic,Zhu2014a,Zhao2019d,guo2022b}. One can construct thermal emitters with nonreciprocal materials featuring an asymmetric dielectric tensor. A conventional way to achieve the nonreciprocal effect is to use the nonreciprocal response of magneto-optical materials under an external magnetic field. The cyclotron motion of electrons breaks the microscopic reversibility~\cite{onsager_reciprocal_1931,onsager_reciprocal_1931-1}, resulting in different properties when the medium is absorbing and emitting light~\cite{madelung2004semiconductors,tanner2019optical}. Using this strategy, Zhu and Fan \cite{Linxiaononreciprocal} proposed a photonic thermal emitter that can achieve complete violation of Kirchhoff’s law. However, since the nonreciprocal effect is quite weak in the thermal radiation wavelength range ($\sim \SI{10}{\micro m}$ at room temperature), the nonreciprocal emitters require a very large external magnetic field of \SI{3}{T} to operate. Later, Zhao et al. \cite{Zhao2019d} improved the photonic design and reduced the required magnetic field to \SI{0.3}{T} at the expense of reduced bandwidth of the nonreciprocal effect. 

\begin{figure}[htbp]
    \centering
    \includegraphics[width=0.85\textwidth]{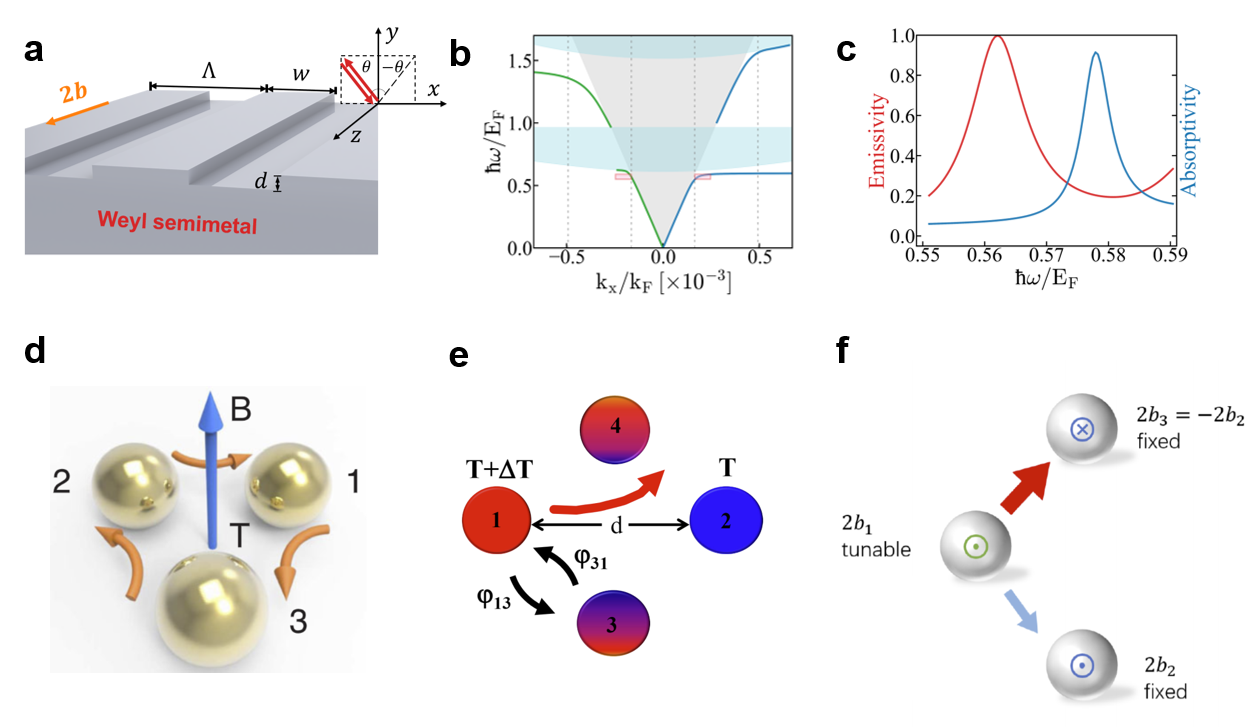}
    \caption{\csentence{Thermal photonic applications.} (a) A nonreciprocal thermal emitter based on a magnetic Weyl semimetal photonic crystal. (b) Dispersion of the nonreciprocal surface plasmon polaritons of the Weyl semimetal. The gray region is the light cone of the vacuum. The blue region denotes the continuum of bulk modes. The red region denotes the surface plasmons used for nonreciprocal thermal emission. The gray dashed lines denote the boundaries of the Brillouin zones. $k_F = E_F/\hbar v_F$ is the Fermi wavevector. (c) Emissivity and absorptivity spectra in $\theta = 80^\circ$ direction. (d) Persistent directional heat current in a many-body system, at the thermal equilibrium of temperature $T$. The spheres are made of magneto-optical materials, with a magnetic field applied perpendicular to the plane of the triangle. Magnetic Weyl semimetals can provide the same effects without a magnetic field. (e) The photon thermal Hall effect among four particles made of magnetic Weyl semimetals forming a square with $C_4$ symmetry. A temperature gradient $\Delta T$ between particles 1 and 2 along the $x$-axis induces a nonzero temperature difference between particles 3 and 4. 
    (f) A radiative thermal router based on three spheres of Weyl semimetal nanoparticles: tuning the Weyl node separation of the center particle will guide the heat to flow in a different direction. Figures are reproduced with permission from (a-c)  Ref.~\cite{zhao_axion-field-enabled_2020},  Copyright 2020 American Chemical Society; (d) Ref.~\cite{Linxiaopersist}, Copyright 2016 American Physical Society; (e) Ref.~\cite{ThermalHall}, Copyright 2020 American Physical Society;
    (f) Ref.~\cite{GuoWeylrouter}, Copyright 2020 American Chemical Society.}
    \label{weyl_thermal}
\end{figure}

As we mentioned in Sec.~\ref{subsec:optical_nonreciprocity}, magnetic Weyl semimetals can exhibit much stronger nonreciprocal effects compared to conventional magneto-optical materials. Such giant nonreciprocity persists at the temperature below the Curie temperature ($T_C$) of the material~\cite{coey2010}. Some of the known magnetic Weyl semimetals possess high Curie temperature; for example, Co$_2$MnGa has $T_C = \SI{694}{K}$~\cite{coey2010}, and Ti$_2$MnAl has $T_C > 650 K$~\cite{feng2015,shi2018b}. These materials are ideal candidates for nonreciprocal thermal emission. In Ref.~\cite{zhao_axion-field-enabled_2020}, Zhao et al.~proposed a broadband thermal emitter based on magnetic Weyl semimetals that achieves near-complete violation of Kirchhoff’s law at room temperatures. The structure is shown in Fig. \ref{weyl_thermal}a which supports nonreciprocal surface plasmon polaritons as shown in Fig. \ref{weyl_thermal}b. The emissivity and absorptivity of the emitter show a significant difference as shown in Fig. \ref{weyl_thermal}c. Tsurimaki et al.~investigated the effect of the Fermi-arc surface state and the number of Weyl nodes for Weyl semimetal-based nonreciprocal thermal emitters~\cite{ChenWeyl1,ChenWeyl2}. Wu et al.~construct nonreciprocal thermal emitters using Weyl semimetal thin films~\cite{wu2021c,wu2021d,wu2022}. Since the nonreciprocal effect is intrinsic to the materials, these thermal emitters do not require any external magnetic field. 

Here we note that for a nonreciprocal thermal emitter, the emissivity and absorptivity in the same direction can be different; however, the angular distribution of emissivity and absorptivity can be still constrained by compound symmetries as revealed in Ref.~\cite{guo2022b}. The compound symmetry can be used to design nonreciprocal thermal emitters with correlated patterns of emissivity and absorptivity~\cite{park_violating_2021}. Breaking the compound symmetries can remove such correlation constraints~\cite{PhysRevApplied.16.064001,guo2022b}.

The above-mentioned works primarily focus on linear polarized thermal radiation. It has been shown that unpatterned thermal emitters with nonreciprocal responses could radiate thermal photons that carry a net spin angular momentum~\cite{PhysRevLett.123.055901,PhysRevB.100.081408,Wang:21}. Khandekar et al.~studied spin-resolved Kirchhoff's laws in nonreciprocal systems~\cite{Chinmayfarfield}. Since these thermal photons can carry nonzero spin angular momentum, the emission of such photons can result in a back-action torque on the thermal emitter~\cite{PhysRevLett.123.055901,circularheatandmomentum}. Maghrebi et al.~discussed this fluctuation-driven torque for a topological insulator thin film out of thermal equilibrium with a cold environment~\cite{PhysRevLett.123.055901}. Guo and Fan proposed a single-particle heat engine utilizing this effect~\cite{guo2020single}.

\subsection{Heat flux control}\label{subsec:heat_flux_control}

Radiative heat transfer plays an important role in photon-based energy conversion and thermal management systems \cite{zhang2007,fan2017}. In traditional thermal systems, the radiative heat flux is driven by a nonzero temperature gradient, and the radiative heat transfer is reciprocal, indicating that when the temperatures of two bodies are exchanged, the magnitude of the heat flux transferred between them is unchanged. However, in systems that involve nonreciprocal materials, photons can flow under zero temperature gradient, and radiative heat transfer could be nonreciprocal. 

In Ref.~\cite{Linxiaopersist}, Zhu and Fan reported the existence of a persistent photon heat current in a nonreciprocal multi-body system even without a temperature gradient. As shown in Fig.~\ref{weyl_thermal}d, the arrows indicate the net heat flux direction when the three magneto-optical spheres are at the same temperature. Later, the persistent heat current feature was also reported within a single nonreciprocal body or cavity~\cite{Spinofheatflow,circularheatandmomentum,chinmayNFcirculation}. This phenomenon is closely related to the thermal Hall effect~\cite{ChenHallandPersistant,Chengmagneticgroup}. In Ref.~\cite{photonthermalhall}, Ben-Abdallah reported that with an external magnetic field, nonzero heat flux is induced in the direction that is perpendicular to the temperature gradient between magneto-optical spheres, as illustrated in Fig.~\ref{weyl_thermal}e. In Ref.~\cite{ThermalHall}, Ott et al.~reported the same phenomenon in a multi-body system made of magnetic Weyl semimetals without the need for an external magnetic field. 

Optical nonreciprocity can also enable efficient heat rectification~\cite{otey2010}, as reported in Refs.~\cite{surfacewavediode,thermalrectification,ThermalrectificationBiehs}. These thermal diodes rely on the nonreciprocal optical surface modes or bulk modes that induce an asymmetric thermal resistance. When the temperature gradient alters direction, the magnitude of the thermal resistance also changes. 

Magnetic Weyl semimetals also provide new opportunities for controlling near-field radiative heat transfer due to their anisotropic optical properties and flexible tunability. In Ref.~\cite{WeyltwistNF}, Tang et al.~reported a twist-induced near-field heat modulator based on magnetic Weyl semimetals. The radiative resistance of the system experiences a significant change depending on the relative alignment of the directions of Weyl nodes separation in the thermal emitter and receiver. Other properties of Weyl semimetals such as the Fermi level, the number of Weyl nodes, the Weyl node separations~\cite{JQSRTWeylNF}, and external magnetic field~\cite{tunablemagnetoresistance,GuoWeylrouter} can also be used to 
modulate near-field heat transfer. Using these effects, in Ref.~\cite{GuoWeylrouter}, Guo demonstrated a radiative thermal router based on Weyl semimetals as illustrated in Fig.~\ref{weyl_thermal}f. We also note several recent reports concerning the general properties of these radiative heat transfer systems involving nonreciprocal bodies~\cite{Linxiaomanybody,LinglingPlanar,ChenHallandPersistant,Chengmagneticgroup,TsuChen}.

\subsection{Casimir force control}\label{subsec:Casimir_force_control}

Besides the exciting opportunities in heat transfer, magnetic Weyl semimetals also provide an effective way to control Casimir forces~\cite{RepulsiveCasimir,ChiralCasimir,torqueWeyl,IandIIWeylCasimir}, especially to create repulsive Casimir force, a long sought-after effect~\cite{PhysRevA.9.2078,PhysRevLett.109.236806,PhysRevLett.106.020403}. In Ref.~\cite{RepulsiveCasimir}, Wilson et al.~reported repulsive Casimir force between semi-infinite magnetic Weyl semimetals when the separation of the two semimetals $d \leq 4 c / \sigma_{x y}$, where $c$ is the speed of light in the vacuum, and  $\sigma_{x y}$ is the bulk hall conductivity of Weyl semimetals. Recently, it was shown that the equilibrium Casimir force in nonreciprocal systems can be used for propulsion~\cite{PhysRevLett.126.170401}. We expect that an enhanced propulsion effect can be realized in magnetic Weyl semimetals.

\section*{Conclusion}\label{sec:conclusion}

In conclusion, we have provided an introductory review of Weyl semimetals in photonics. We covered the basic concept and optical properties of Weyl semimetals, and surveyed their emerging applications in photonic science and engineering. We discussed how the nontrivial topology of Weyl semimetals leads to unusual optical properties. Photonics based on Weyl semimetals is an emerging topic with many open challenges and new opportunities. We believe that more exciting applications of Weyl semimetals will come up in the near future, and we wish this pedagogical review will benefit upcoming researchers exploring this new direction.


\begin{backmatter}

\section*{Declarations}

\subsection*{Availability of data and materials}
Not applicable.

\subsection*{Competing interests}
The authors declare that they have no competing interests.

\subsection*{Funding}
This work is supported by MURI projects from the U.~S.~Army Research Office (Grant No. W911NF-19-1-0279) and the U.~S.~Air Force Office of Scientific Research (FA9550-21-1-0244).

\subsection*{Author's contributions}
CG, VSA, BZ, and SF~wrote the manuscript. CG~led the collaboration. SF~supervised the work. 

\subsection*{Acknowledgements}
Not applicable.


\bibliographystyle{bmc-mathphys-no_urldate} 
\bibliography{bmc_article}      

\end{backmatter}
\end{document}